\documentclass{aa}  
\usepackage{natbib}
\bibpunct{(}{)}{;}{a}{}{,}
\usepackage{graphicx}
\begin{document}

\authorrunning{S. Criscuoli, P. Romano et al. }
\titlerunning{Magnetic evolution of superactive regions}
   \title{Magnetic evolution of superactive regions}
\subtitle{Complexity and potentially unstable magnetic discontinuities}
   \author{S. Criscuoli \inst{1}, P.Romano \inst{2}, F. Giorgi \inst{1}, F. Zuccarello \inst{3}}

   \offprints{Serena Criscuoli \email{criscuoli@oaroma.inaf.it}}

   \institute{ INAF - Osservatorio Astronomico di Roma, Via Frascati 33, I-00040, Monte Porzio Catone, Italia
	\and
INAF- Osservatorio Astrofisico di Catania, via S. Sofia 78, 95123 Catania, Italy
   \and
Dipartimento di Fisica e Astronomia - Sezione Astrofisica, Universit\'a di Catania, via S. Sofia 78, 95123 Catania, Italy
   }
   \date{}

   \abstract
{It is widely accepted that solar flares are manifestations of magnetic reconnection events taking place in the solar atmosphere. Several aspects of these events remain unclear, although many efforts have been devoted to the investigation of magnetic field configurations at flare occurrence sites.} 
{In this work, we have studied the temporal evolution of some properties of a sample of superactive regions with the aim to single out the most significant for flare activity forecasting.}
{We have investigated properties of 14 superactive regions, observed between January 1st 2000 and December 31st 2006 with MDI/SOHO instrument and characterized by a particularly intense flare activity during their passage on the solar disk. We have analyzed the temporal evolution of fractal and multifractal properties of photospheric magnetic fields, namely the generalized fractal dimension and the cntribution and dimensionality diversities, which describe geometrical properties of the magnetic field, as well as the potential unstable volumes of magnetic discontinuities above the studied ARs, which may provide information about the magnetic field configuration in upper layers of the atmosphere. Correlations of these quantities with the flare index, which provides information about the flare activity of a region, have also been estimated.}
{We found that in 50 \% of our sample the generalized fractal dimension is correlated with the flare index computed over windows of 50 hours, while the contribution diversity and the dimensional diversity are anticorrelated with the same index. A clear increase of the potential unstable volume of magnetic discontinuities in the corona is observed before the phases characterized by more frequent and intense flares. We also found that the free energy distribution functions of unstable volumes of the analyzed superactive regions can be fitted with straight lines whose slope is larger than the values found in previous works for less active magnetic regions.}
{The generalized fractal dimension and the potential unstable volume of magnetic discontinuities are the most suitable for statistical investigations of relations with flare activity over longer (50 hours) and shorter (few hours) time intervals, respectively.}


\maketitle

\section{Introduction}
     \label{S-Introduction}

It is largely accepted that solar flares are magnetic reconnection phenomena that can occur at different scales of time and energy \citep {Asch00}. Although the flare intensity and frequency are proportional to the magnetic field strength involved in these events, it is still unclear what the magnetic configurations are that drive these energy releases. In the past decade the theory and the numerical simulations of magnetic reconnection have shown that the dissipation of magnetic energy, responsible for the flare activity, occurs along unstable magnetic discontinuities \citep {Parker1988}. The formation and disappearance of these discontinuities, also named current sheets, is a continuous process which occurs in all magnetic concentrations as in the solar active regions. Due to the thickness of these discontinuities, below current telescope resolution, many authors have tried to deduce their presence and location from the global topology of active regions.

Many statistical studies have been devoted to the investigation of physical properties and topology of the magnetic field in the photosphere, compatible with the events \citep[see] [ and references therein]{schrijver2007,schrijver2009}, but the results are controversial \citep{georgoulis2007}. \citet{romano2007} (Paper I, hereafter) investigated the correlation of flare productivity and some properties of 26 superactive regions.
They found that the deviations from magnetic polarity balance
and the length of the inversion lines characterized by a horizontal magnetic gradient greater than 0.3 G km$^{-1}$ seem to have a major role in producing a magnetic field topology able to trigger flares.  This result has been confirmed by \citet{song2009}, who analyzed 174 active regions and employed a ordinal logistic regression model to predict solar flares. 
Different conclusions were drawn by \citet{leka2007}, who applied discriminant analyses to almost thirty properties of the photospheric magnetic field estimated for 496 active regions. Their results indicate that {\em available data do not provide a basis for reliable flare prediction.} However, these authors also suggest that different results might be found by including  descriptors of the self-similar properties of the photospheric magnetic field in the analyses.

Self-similar properties of the solar magnetic field are a signature of the turbulent physical processes which determine the generation, evolution and dissipation of magnetic structures \citep[see for instance][ and references therein]{cattaneo2003,abramenko2005c, stein2006} and have been investigated in several studies \citep[see for instance][ for a review]{georgoulis2005}.
In this context, investigation of fractal and multifractal properties of photospheric magnetic features provide important information about the physical properties of convective plasma \citep{balke1993,tao1995,nesme1996}, of the magnetic field \citep{stenflo2002,pietgraham2009} and about the physical processes that govern their interaction \citep{lawrence1995,lawrence1999,bushby2005,crouch2007}. Studies carried out on short temporal ranges (hours-days), have showed that fractal and multifractal properties vary with the evolution of the photospheric magnetic field of the regions analyzed \citep[e.g.][]{georgoulis2005,abramenko2005,conlon2008,hewett2008}. Several authors have also investigated the self-similar properties of ARs in connection with their flare productivity.  For instance \citet{abramenko2005b} found that the magnetic power spectrum of 16 investigated ARs becomes steeper with the increase of the flare productivity. \citet{abramenko2005} also compared the multifractal properties of two regions characterized by different flare activity levels and found that the more flare productive region showed a higher level of multifractality. A similar investigation was carried out by \citet{georgoulis2005}, who also presented the multifractal analyses as more suitable than the fractal one for flare forecasting. Statistical analysis of large samples of data have shown that variations of fractal properties of ARs in photosphere  \citep{meunier2004} and low chromosphere \citep{criscuoli2007} do not present clear trends with the activity cycle. On the contrary, \citet{mcateer2005} reported a correlation between variations of  photospheric fractal properties of ARs with flare productivity. 

Precursors to flares have also been looked for in the solar corona, where the sites of magnetic reconnection are usually located. Variations of magnetic field characteristics and topology in connection with occurrence of eruptive phenomena have been investigated  \citep[e.g.][]{Ugarte2007,schrijver2009}. Among the most recent studies, \citet{barnes2006} applied discriminant analyses to several properties of coronal fields above three active regions and concluded that coronal field descriptors are more efficient than photospheric ones to distinguish between flaring and non-flaring regions. 
Self-similarity is also a characteristic of coronal magnetic fields  \citep[e.g.][ and references therein]{georgoulis2005, belanger2007}. \citet{aschwanden2008I,aschwanden2008II} measured the fractal dimension of flare arcades and found it to be linearly correlated with the flare flux in EUV. \citet{abramenko2008} estimated the intermittency of the magnetic field of an AR and found a time delay of about one day between the evolution of intermittency in photosphere and in corona. \citet{vlahos2004} investigated properties of the magnetic discontinuities of the solar corona, i.e. regions characterized by magnetic field variations occurring on short scale lengths and where the nonpotential magnetic energy is accumulated. While most of these discontinuities remain relatively stable, a fraction of them may become unstable, thus causing rapid magnetic reconnection that releases the free magnetic energy and provides the energy content for flaring events. \citet{vlahos2004} studied an active region characterized by no flare or subflare occurrence and an active region showing a number of subflares but no major flares over the course of several hours. They used linear force-free field extrapolations to calculate the 3D magnetic configuration and various selection criteria to identify these discontinuities. They found that the free magnetic energies and volumes contained in these discontinuities obey power-law distribution functions, with power-law indices for the free energies in the range [-1.6, -1.35].

In this paper, we extend the work presented in Paper I by investigating other properties of photospheric and coronal magnetic fields of the same sample of superactive regions. We consider the temporal evolution of some fractal and multifractal properties of the photospheric field, namely the generalized, the contribution and the diversity dimensions \citep{conlon2008} (see \S3.1 for the definition of these parameters), estimated by the analyses of MDI magnetograms. We also investigate the evolution of the free energy distribution and the potentially unstable volume of the magnetic discontinuities \citep{vlahos2004} of the coronal field  (see \S3.2) as derived by force-free field extrapolation. We investigate the temporal correlation with the flare activity in order to single out the indicators that best describe the flare productivity potential of a region.
We choose these indicators because they describe the topological complexity of the magnetic field in different layers of the solar atmosphere. 

This paper is organized as follows. The data-set analyzed and the indicators employed for this study are described in \S2 and \S3, respectively. In \S4 we present the results we obtained and in \S5 we discuss them and draw our conclusions.

\section{Data}
\label{Data}

According to criteria explained in Paper I, we selected 26 superactive regions (ARs) which appeared on the solar disk during the middle of the 23th solar cycle, between January 1st 2000 and December 31st 2006 and characterized by a flare index \citep{li2004} greater than 500, computed over time intervals of 50 hours. The X-ray flare index I computed over the time interval $\Delta t$ is defined as:
\begin{eqnarray}
I(\Delta t) &=&0.1\times\sum_h{B_h}+\sum_i{C_i}+10\times\sum_j{M_j}+ \nonumber\\ 
     &+&100\times\sum_k{X_k}
\label{Index}
\end{eqnarray}
where $B_h$, $C_i$, $M_j$ and $X_k$ refer to the soft X-ray classification based on
the flux in the 1-8 $\AA$ range of the spectrum \citep{tandberg1988}. Depending on the X-ray flux, flares are classified as B, C, M or X,
where B are the less energetic flares ($10^{-4} erg \cdot cm^{-2}\cdot s^{-1}$) and X are the
most energetic ($10^{-1} erg \cdot cm^{-2}\cdot s^{-1}$). The sums in eqn.\ref{Index} run over the number of flares of that class produced in an active region in the time interval $\Delta t$.  

For each region we used MDI/SOHO line-of-sight full-disk magnetograms taken during the time interval between the first and the last flare, with a time resolution of 96 minutes and a plate scale of 1.98 arcsec/pixel. We considered only those magnetograms where the longitudinal angular distance of the active region from disk center was less than 41 degrees. Among these 26 ARs we neglected those for which MDI showed a saturation effect in sunspot umbra \citep {liu2007}, thus reducing the analysis to 14 ARs. 
\begin{table*}[ht]
\caption{Superactive regions selected, correlation coefficients between the investigated indicators and the flare index, average value of indicator $\alpha_E$.} 
\centering
\begin{tabular}{c c c c c c c c c}
\hline
 \hline
NOAA & First Flare & Last Flare & $N_{mag}$ & $D_{8}$ & $D_{d}$ & $C_{d}$ & $I_{v}$ & $<\alpha_{E}>$\\
\hline
9026* & Jun. 1,2000 01:01 UT & Jun. 10,2000 178:19UT & 48 &  0.68 &  -0.62  & -0.34 & -0.38 & -1.6 $\pm$ 0.2\\
9236 & Nov. 22,2000 04:59 UT &  Nov. 27,2000 00:21UT & 82 &  0.75 & -0.68 & -0.59 & 0.54 & -1.7 $\pm$ 0.3\\
9690 & Nov. 6, 2001 08:34 UT & Nov. 15, 2001 17:29UT & 84 & 0.66 & -0.77 & -0.65 & 0.55 & -1.7 $\pm$ 0.4\\
9733 & Dec, 8, 2001 02:56 UT & Dec. 19, 2001 01:24 UT & 87 & -0.72 & 0.62 & 0.1 & 0.44 & -1.55$\pm$ 0.4\\
10030 & Jul. 10, 2002 01:56 UT & Jul. 20, 2002 12:44 UT & 94 & -0.02 & 0.04 & 0.29 & -0.40 & -1.8 $\pm$ 0.4\\
10039* & Jul. 23, 2002 00:47 UT & Aug. 3, 2002 19:11 UT & 33 & -0.02& 0.54 & 0.66& 0.04 & -1.7 $\pm$ 0.2\\
10314 & Mar. 15, 2003 03:44 UT & Mar. 22, 2003 23:03 UT & 46 & 0.23 & -0.57 & -0.38 & 0.04 &  -1.7 $\pm$ 0.2\\
10365* & May 25, 2003 20:38 UT & Jun. 2, 2003 16:01 UT & 68 & 0.81 & -0.78 & -0.66 & -0.29 & -1.7 $\pm$ 0.3\\
10375* & Jun. 1, 2003 04:00 UT & Jun. 15, 2003 12:35 UT & 83 & 0.5& -0.03 & 0.14 & 0.46 & -1.65$\pm$ 0.3\\
10486* & Oct. 22, 2003 05:21 UT & Nov. 5, 2003 10:56 UT & 80 & -0.2& -0.03& 0.42 & -0.93 & -1.45$\pm$ 0.2\\
10656* & Aug. 6, 2004 23:01 UT & Aug. 19, 2004 14:16 UT & 76 & 0.88& -0.66& -0.59 & 0.84 & -1.7 $\pm$ 0.3\\
10696 & Nov. 2, 2004 16:42 UT & Nov. 11, 2004 16:02 UT & 91 & -0.11& 0.68& 0.54 & 0.22 & -1.7 $\pm$ 0.2\\
10720 & Jan. 11, 2005 20:46 UT & Jan. 23, 2005 05:23 UT & 81 & -0.03& 0.1& 0.06 & 0.38 & -1.8 $\pm$ 0.3\\
10786* & Jul. 1, 2005 05:07 UT & Jul. 15, 2005 17:15 UT & 74 & 0.5& -0.35& -0.41 & -0.77 & -1.8 $\pm$ 0.2\\
\hline
\end{tabular}
\label{tab1}
\end{table*}
For each region we considered a sub-field centered on the baricenter of the AR such that the maximum magnetic flux imbalance was less than 20$\%$. Sub fields of size smaller than 100$\times$100 pixels were discarded. The number of magnetograms for each AR selected according to these criteria is given in Table \ref{tab1}. Geometrical projection was compensated for by the \textit{ideproject} Solar Soft routine \citep{freeland1998}. 

For 7 ARs (marked with a '*' in Tab. \ref{tab1}), which have been observed by TRACE, we also considered high resolution images at 171 \AA~or 195 \AA~ in order to investigate their coronal configuration.

%______________________________________________________________
\section{Data analysis}

From each line-of-sight magnetogram sub-field we computed 5 parameters, three of which describe fractal and multifractal properties of the photospheric magnetic field and two which describe the behavior of the upper level of the solar atmosphere above the selected ARs. Then we investigated the correlation between these properties and the flare index computed over temporal windows of 50 hours.

\subsection{Fractal and multifractal indicators}

To investigate the fractal properties of ARs, we estimated the \textit{generalized fractal dimension $ D_{8}$} by using the box-counting technique. The computation of the \textit{Contribution Diversity $C_{d}$} and \textit{Dimensional Diversity $ D_{d}$} allowed us to investigate multifractal properties. As already pointed out in \S1, the literature concerning fractal and multifractal studies in solar physics is quite extensive; here we summarize the main points.

The most common method of estimating the fractal dimension of an object given its image is box counting. It consists of covering the image with boxes of different sizes $\epsilon$ and then estimating the slope of the linear relation
\begin{equation}
\label{fracdim0}
	\ln N(\epsilon)=  D\lim_{ \epsilon \to 0} \ln  \frac{1}{\epsilon} + C
\end{equation}
 
where $N(\epsilon)$ is the number of boxes that cover the object, $C$ is a constant and $D$ is the fractal dimension.
More in general, given a measure $P$, the \textit{generalized dimension} is defined as
\begin{equation}
\label{fracdim}
	D_q=\frac{1}{q-1}\lim_{ \epsilon \to 0}\frac{\ln I_{q}(\epsilon)}{\ln \epsilon}
\end{equation}

where $q$ is a real number, 
\begin{equation}
I_{q}(\epsilon)=\sum_{i=1}^{N_{\epsilon}}P_{i}(\epsilon)^{q}
\end{equation}
and we have introduced the more compact notation $N_{\epsilon}$ instead of $N(\epsilon)$. As $P_{i}(\epsilon)$ we consider the normalized magnetic flux $\frac{|\sum_{j}\phi_{j}|}{\phi_{tot}}$, where $j$ runs in a box of size $\epsilon$ and ${\phi_{tot}}$ is the total flux in the image. The variation of the exponent $q$ allows us to investigate different spatial scales: negative values of $q$ enhance the smallest spatial scales whereas positive values enhance the largest spatial scales. $D_{q}$ is a monotonic decreasing function of $q$. The R\`enyi dimension $D_{0}$ is the fractal dimension. Instead of $D_{0}$, we investigated the temporal variation of the generalized dimension $D_{8}$, where, according to equation \ref{fracdim},
\begin{equation}
\label{fracdim8}
	D_8=\frac{1}{7}\lim_{ \epsilon \to 0}\frac{\ln I_8(\epsilon)}{\ln \epsilon}.
\end{equation}

We choose $D_{8}$ since it is less sensitive to spatial resolution and pixelization effects \citep{lawrence1996}. 

The strength of the measure $\alpha_q$ is defined as
\begin{equation}
\label{alfa}
\alpha_{q}=\lim_{\epsilon \to 0}\frac{\sum_{i=1}^{N_{\epsilon}}\eta_{i,q}(\epsilon)\ln P_{i}(\epsilon)}{\ln \epsilon}
\end{equation}

where $\eta_{i,q}(\epsilon)=\frac{P_{i}(\epsilon)^{q}}     {\sum_{i=1}^{N_{\epsilon}}P_{i}(\epsilon)^{q}}$. 

The multifractal spectrum $F(\alpha_q)$ is defined as 
\begin{equation}
F(\alpha_q)=\lim_{\epsilon \to 0}\frac{\sum_{i=1}^{N_{\epsilon}}\eta_{i,q}(\epsilon)\ln\eta_{i,q}(\epsilon)}{\ln \epsilon}.
\end{equation}

It can be shown that, under the assumption $F''(\alpha_q)<0 \Leftrightarrow dD_q/dq<0$, 

\begin{equation}
F(\alpha_q)=q\alpha_q-(q-1)D_{q}  
\end{equation}

$\alpha_q$ is the exponent which describes the scaling of the measure at a certain position of the domain, while $F(\alpha_q)$ is its distribution \citep{halsey1986}. For a bi-dimensional image $0\leq\alpha_q\leq 2$, where $\alpha_q =0$ in the case of a sharp spike and $\alpha_q =2$ for a homogeneous image.
Therefore, low values of $\alpha_q$ indicate denser regions, while high values indicate more rarefied regions, and the value of $\alpha_q$ describes the degree of 'spikiness' of an image \citep{chappell2001}.
The set of points of the domain with the same value of $\alpha_q$ has fractal dimension $F(\alpha_q)$. The spectrum is a concave function of $\alpha_q$ and its maximum is the fractal dimension $D_0$. For a single fractal of fractal dimension $D_f$,  $\alpha_q=F(\alpha_q)=D_f,  \forall q$. 
The contribution diversity $C_{d}$ and the dimensional diversity $D_{d}$ \citep{conlon2008} are defined respectively as 
\begin{equation}
C_{d}=(\alpha_q)_{max}-(\alpha_q)_{min}    ;\\
D_{d}=F(\alpha_q)_{max}-F(\alpha_q)_{min}.  
\end{equation}

Therefore, the contribution diversity  and the dimensional diversity describe the variety of the concentration of the different structures in an image (spikes, filaments, clumpy structures) and of the their complexity. Following \citet{conlon2008}, the analysis was restricted to $q>0$, since numerical errors are larger for negative values of the exponent $q$.
% Fits in \ref{fracdim} and \ref{alfa} are performed in the linear range only \citep{lawrence1996}. 

\subsection{Volume and free magnetic energy of unstable magnetic discontinuities}

 For each magnetogram we performed a linear force-free field extrapolation to determine the 3-D magnetic field configuration. We applied an extrapolation code based on the method introduced by \citet{alissandrakis1981}, which assumes that the magnetic field is force-free not only in the corona but also in the lower levels and that it vanishes at infinity. In order to minimize the oscillatory behavior of the Fourier solutions, we put the original arrays of the data for the boundary in bigger arrays of double dimensions, before applying the extrapolation procedure. A consideration of the size of the integration box gives a resonant value of the force-free parameter. Below this value the periodicity effects inherent to the
Alissandrakis method are reduced. Therefore, for each AR we fixed the force-free parameter just below the resonant value for the magnetogram with the largest field of view among the magnetograms of the same AR. Moreover, in order to investigate the sensitivity of our results to variation of this parameter, for 7 AR we also performed extrapolatons using the force-free parameters which best fit the coronal loops observed at 171 \AA~or 195 \AA with TRACE.

Following the method used by \citet{vlahos2004} we applied a selection criterion based on the Parker angle in order to identify potentially unstable regions inside each AR. In particular, we considered as potentially unstable the location where the angular difference between two adjacent magnetic field vectors was greater than 30$^{o}$. Then, we computed the normalized total volume occupied by all the potentially unstable regions:
\begin{equation}
I_{v}=\frac{\sum_{i=1}^{N}n_{i}\lambda^{2}\delta h}{I_T},
\end{equation}
where N is the number of the potentially unstable regions, $n_{i}$ is the number of adjacent locations inside each region, $\lambda$ is the pixel size of the magnetogram and $\delta$h is the height step of the force-free field extrapolation. The normalization to the total volume of the box $I_T$ in which the extrapolation was performed takes into account the variable sizes of the sub-field magnetograms analyzed. 

We also measured the free magnetic energy $E_{i}$ in each unstable region $i$ consisting of a set of neighbouring grid points $j$ where the instability criterion is met:
\begin{equation}
E_{i}=\frac{\lambda^{2}\delta h}{8\pi}\sum_{j=1}^{n_{i}}B_{ff,j}^{2}-B_{pf,j}^{2},
\end{equation}
with $B_{ff,j}$ and $B_{pf,j}$ the magnetic field values deduced by the linear force-free field and the potential field, respectively. Note that this formula, which has been widely used in the literature \citep[e.g.][]{jing2009,bobra2008,kusano2002}, differs from the one reported in \citet{vlahos2004} for reasons not known to these authors.

\begin{figure*}
\centering{
\includegraphics[width=4.cm]{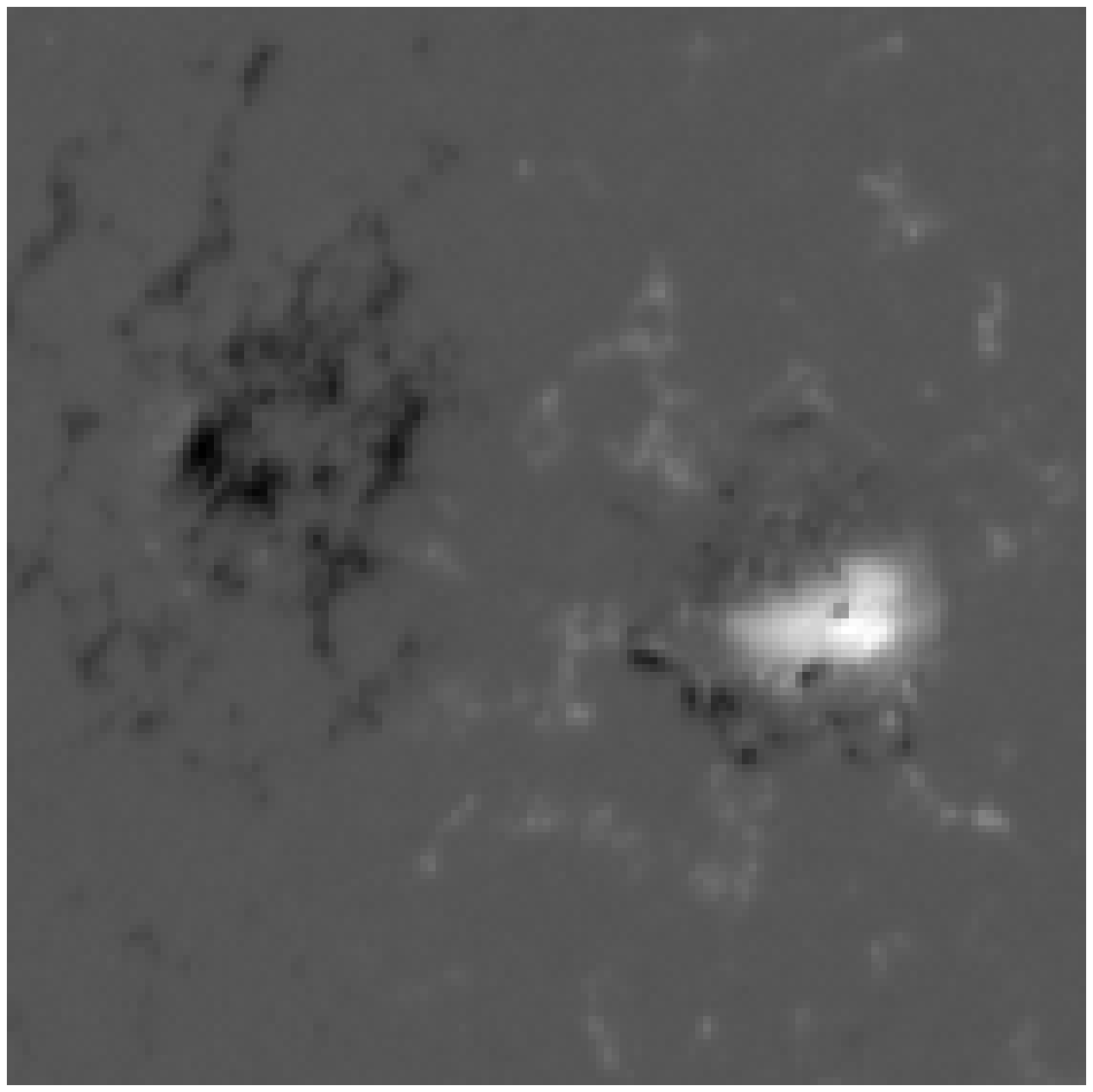}
\includegraphics[width=4.cm]{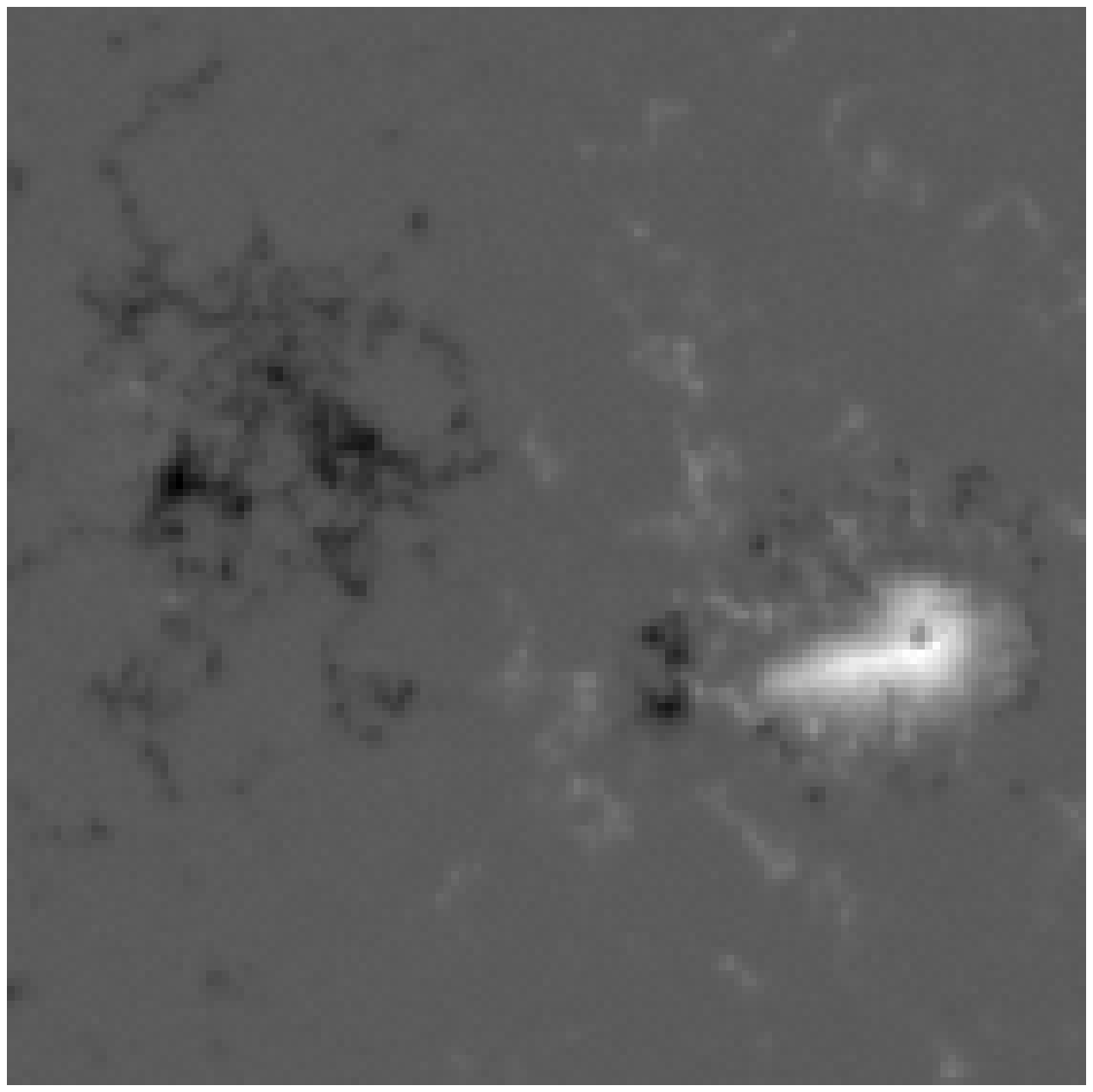}
\includegraphics[width=4.cm]{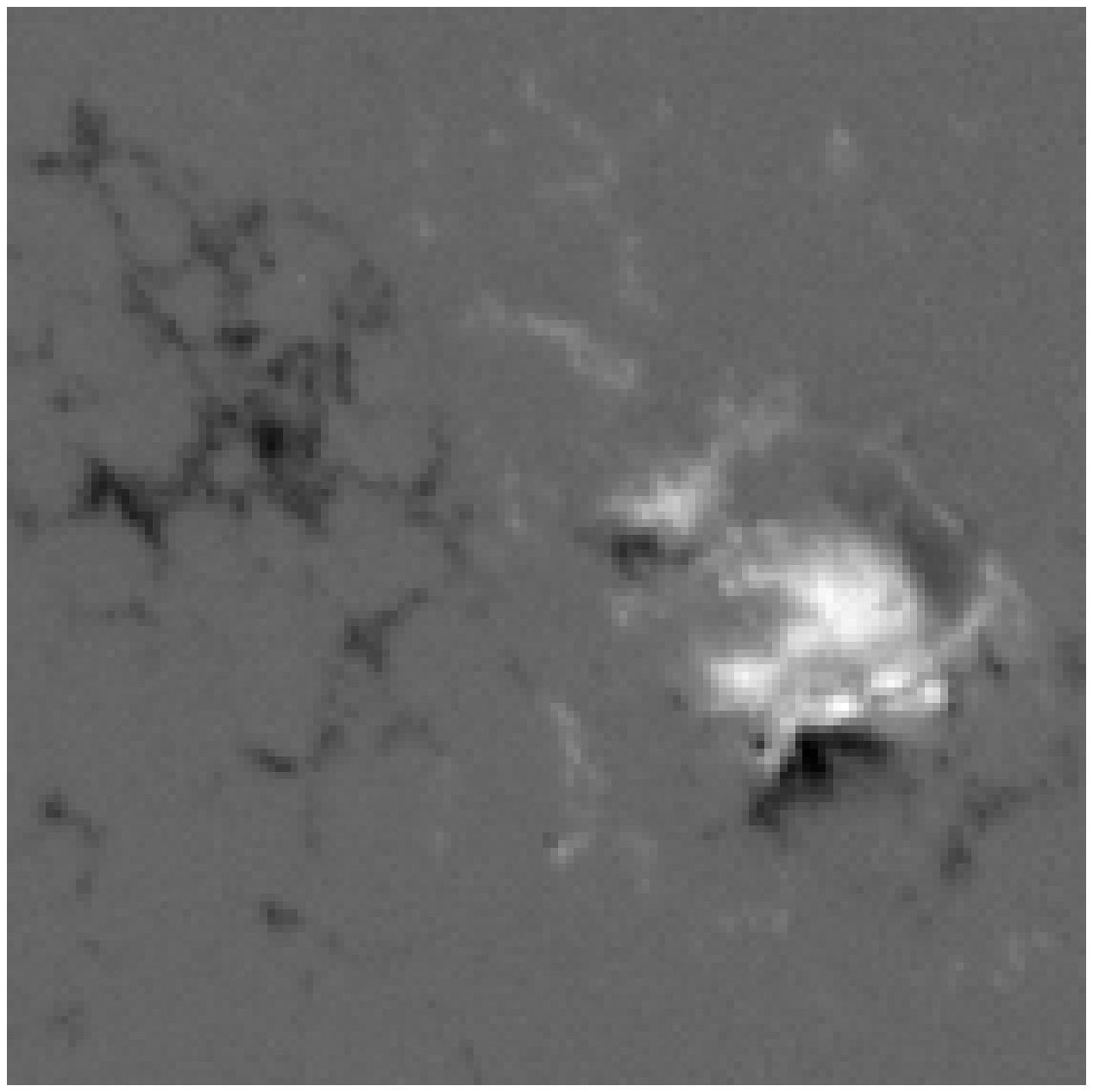}\\
\includegraphics[width=4.cm]{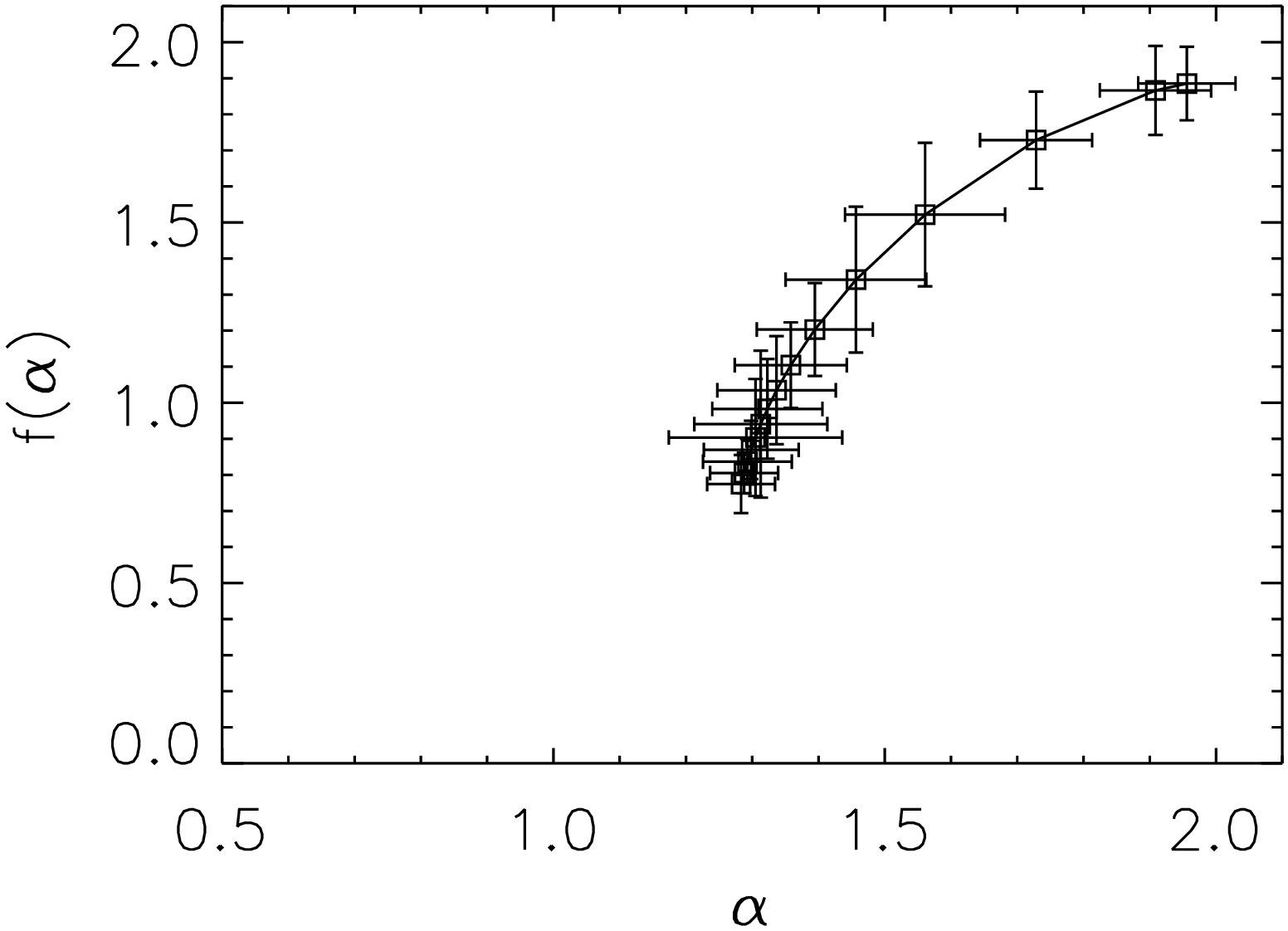}
\includegraphics[width=4.cm]{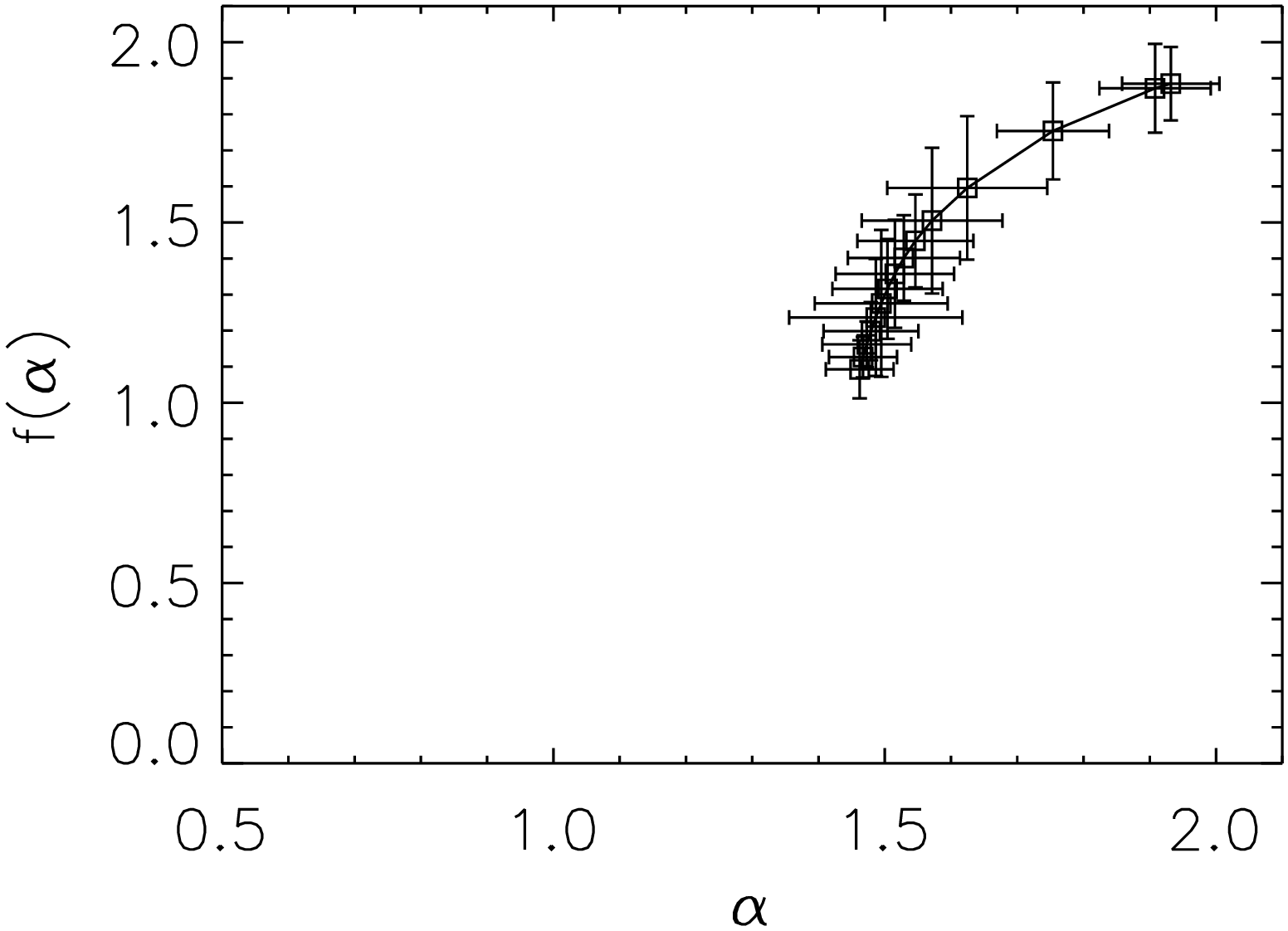}
\includegraphics[width=4.cm]{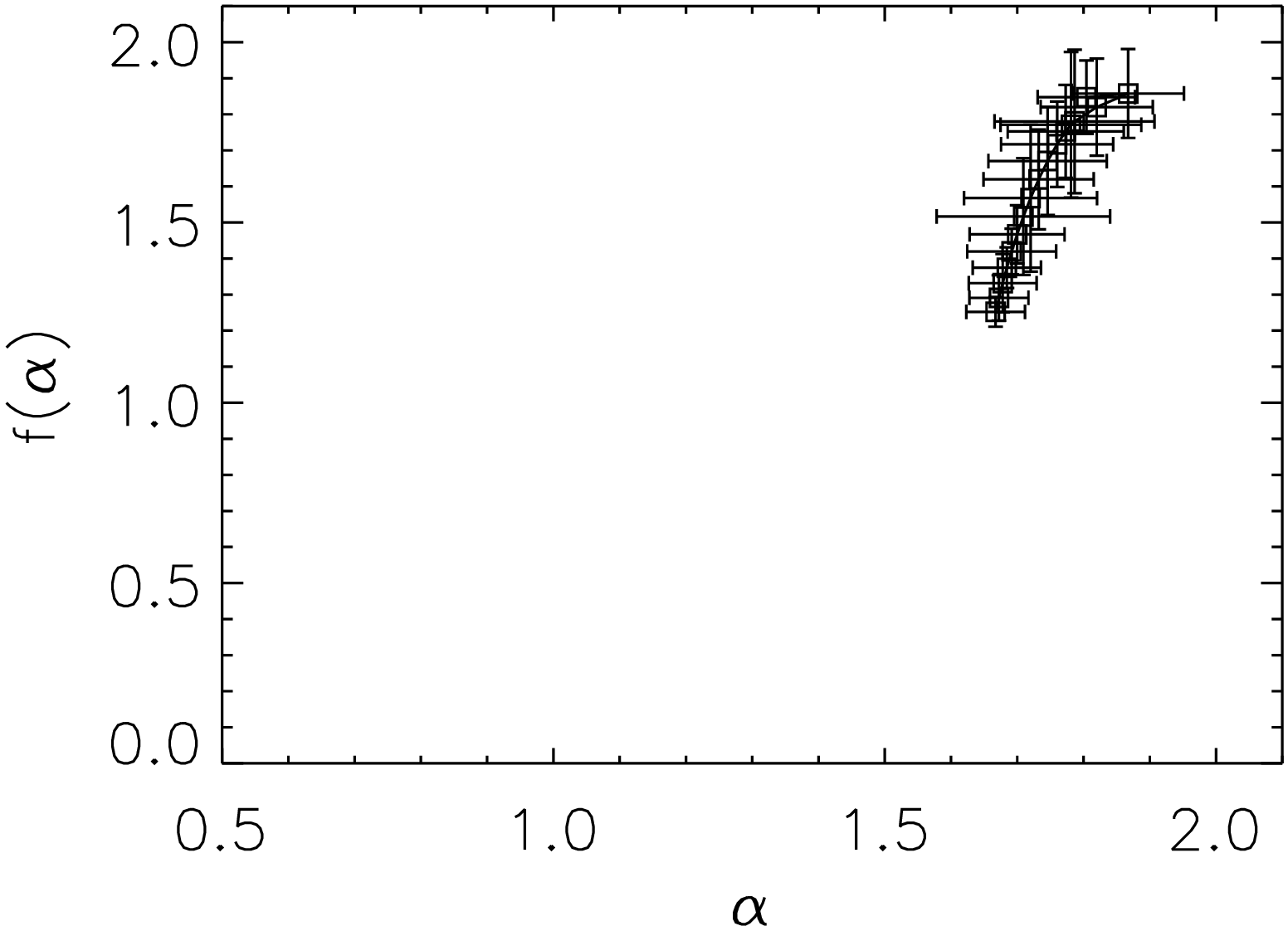}\\
\includegraphics[width=4.cm]{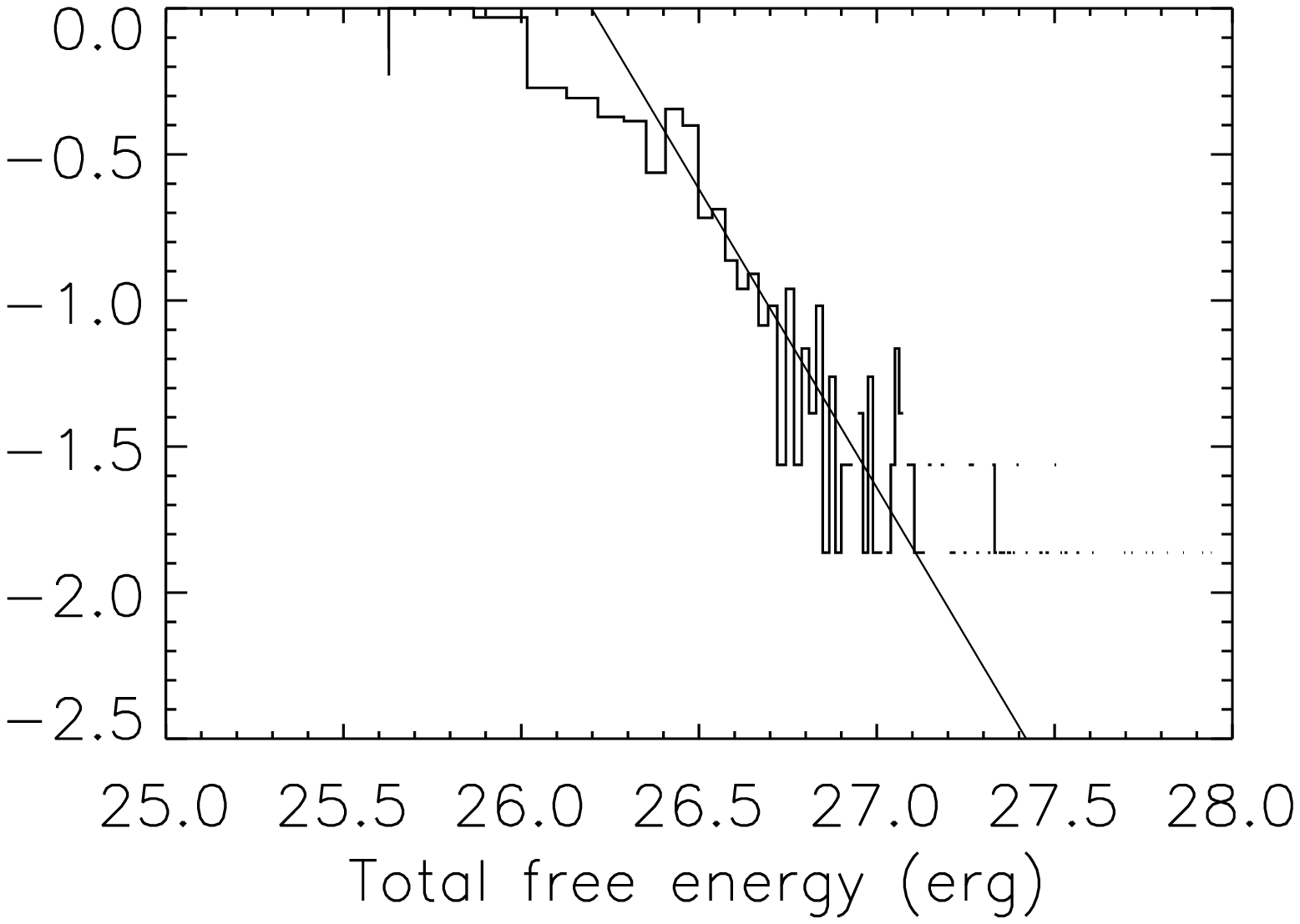}
\includegraphics[width=4.cm]{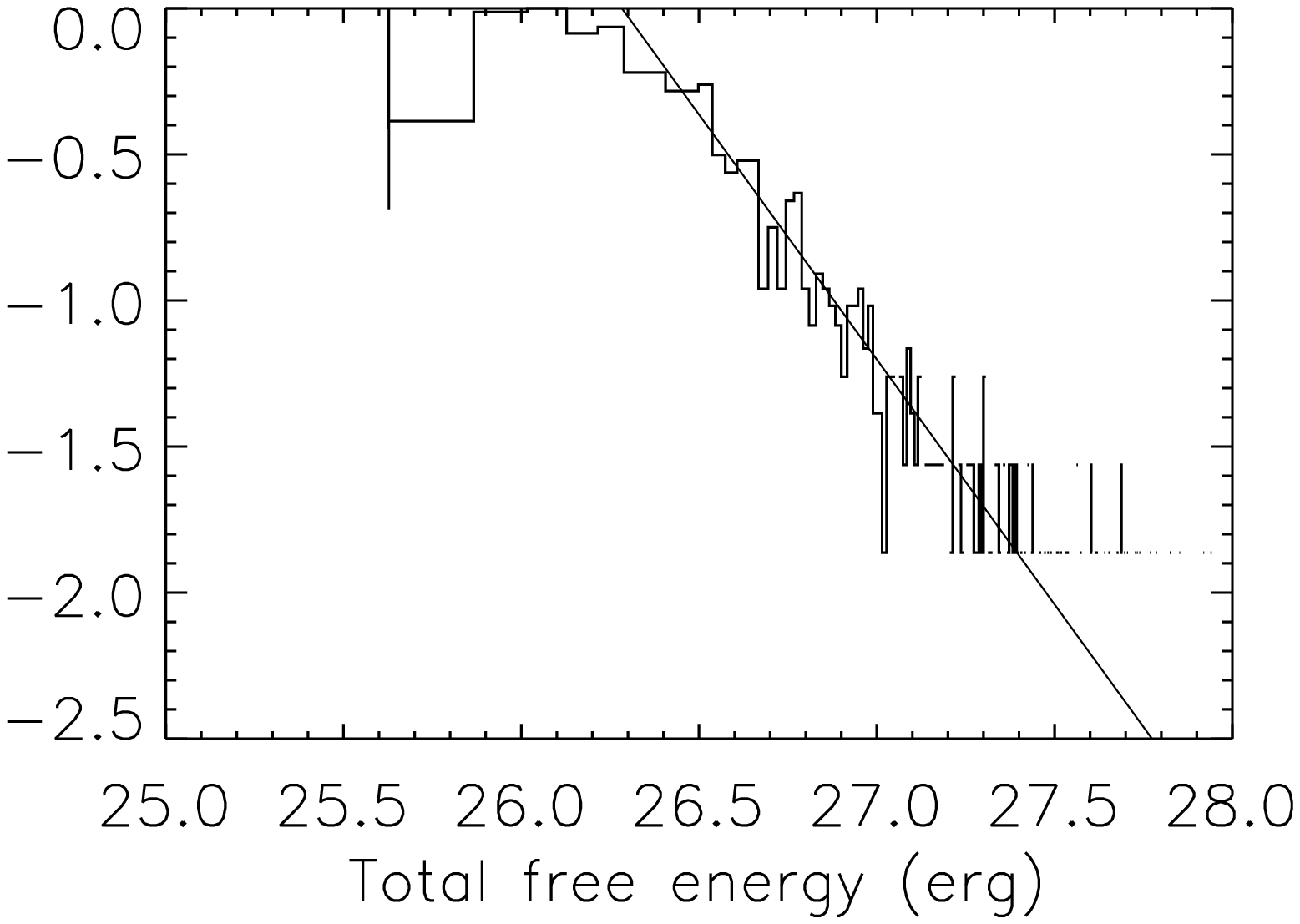}
\includegraphics[width=4.cm]{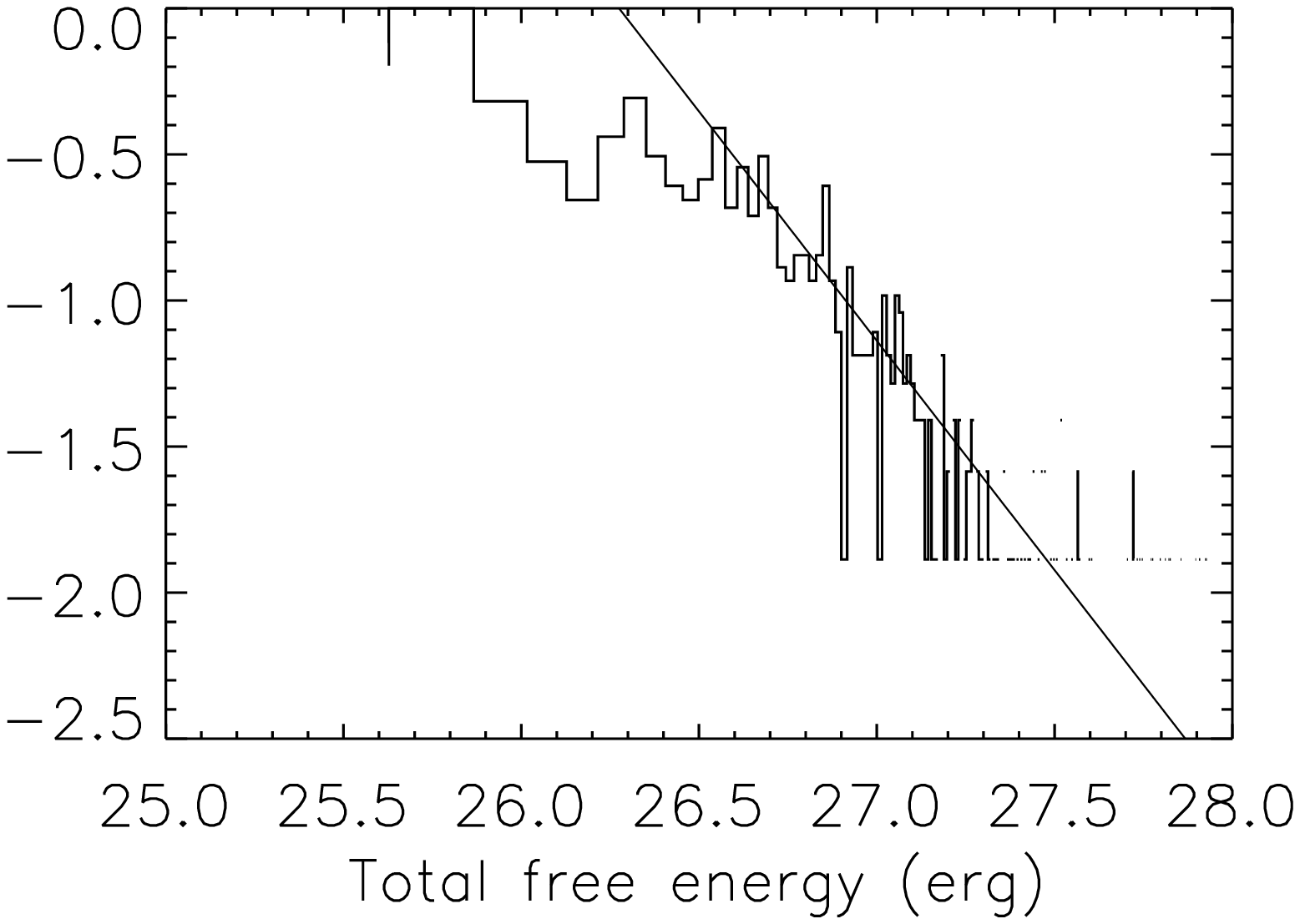}
} 
\caption{Top: Details of  MDI magnetograms, acquired  on 22 November 2000 04:51 UT (left), 23 November 2000 11:20 UT (center), 26 November 2000 19:11 UT (right), showing the emergence phase of NOAA 9236. Middle: corresponding multifractal spectra. Bottom: corresponding distribution functions of total free energy.}
\label{fig_9236}
\end{figure*}

\begin{figure*}
\centering{
\includegraphics[width=4.cm]{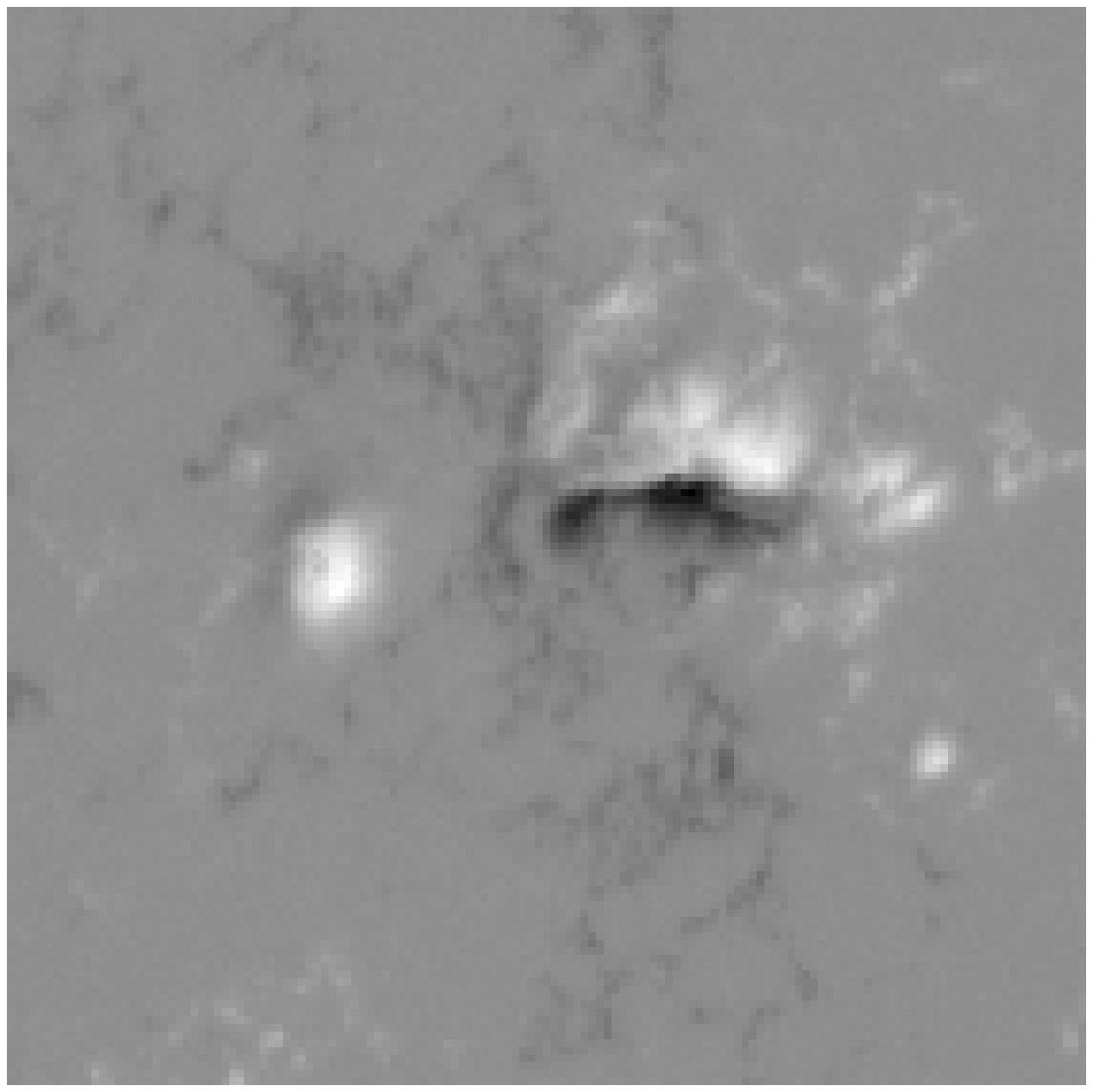}
\includegraphics[width=4.cm]{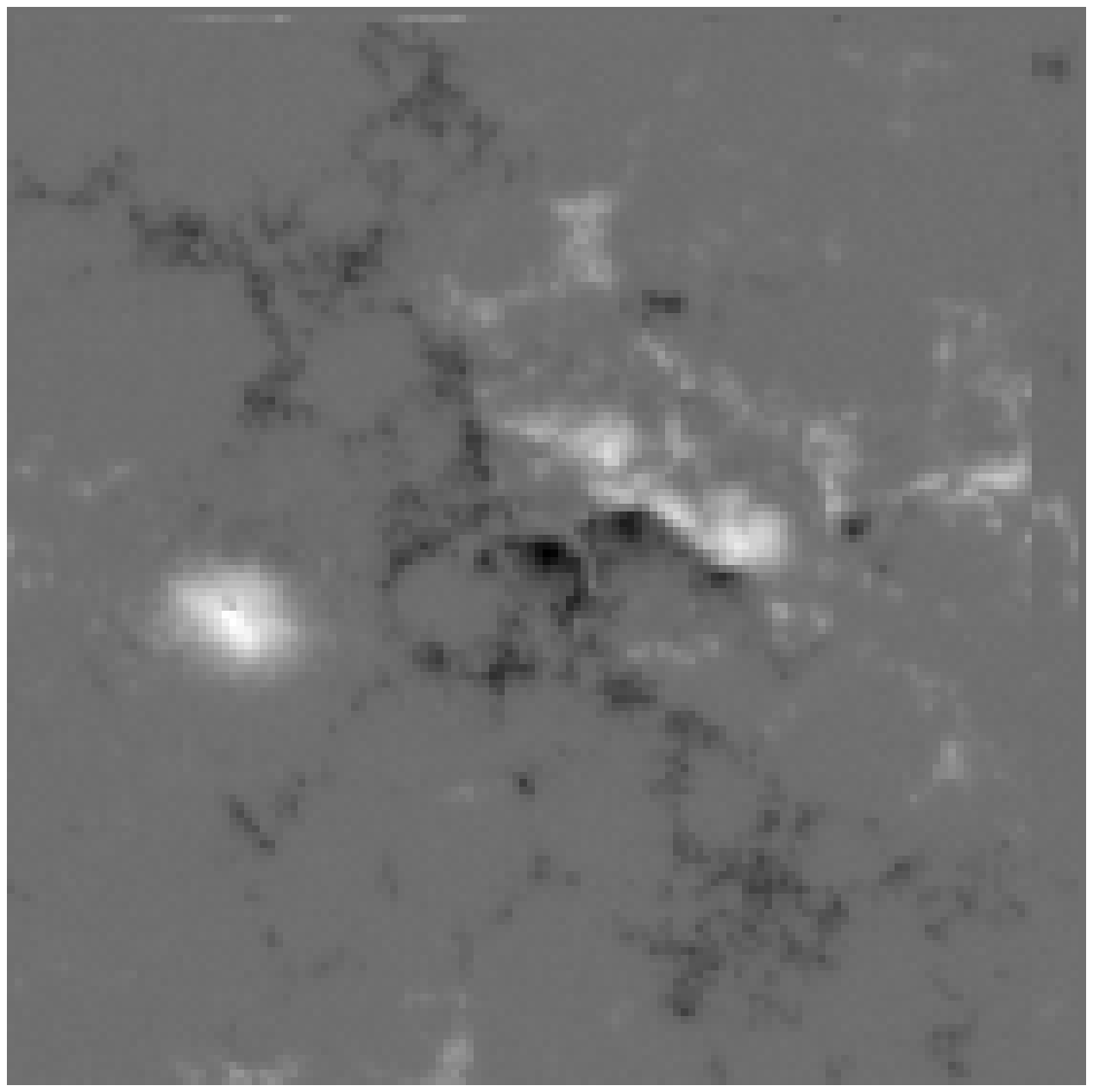}
\includegraphics[width=4.cm]{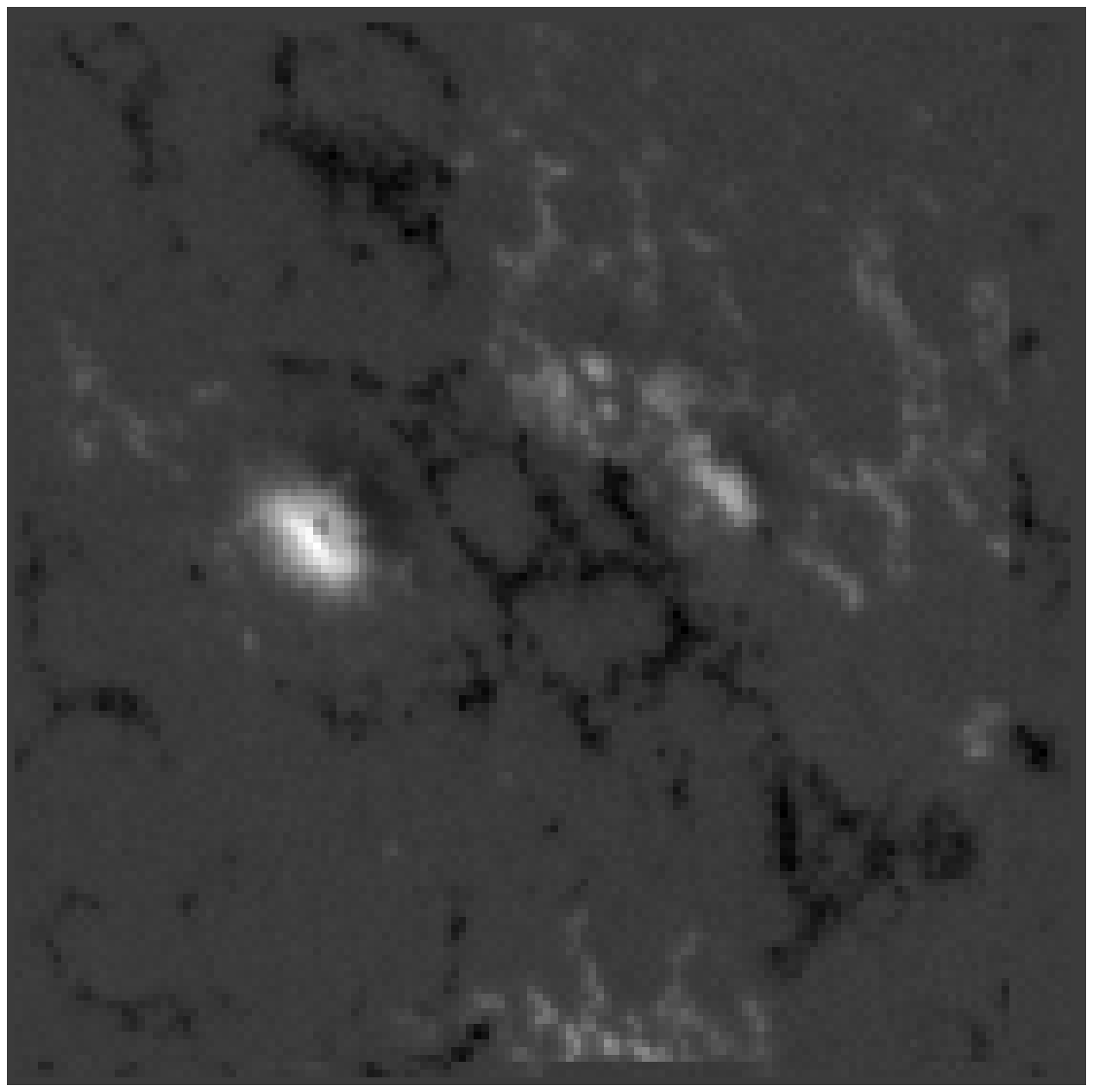}\\
\includegraphics[width=4.cm]{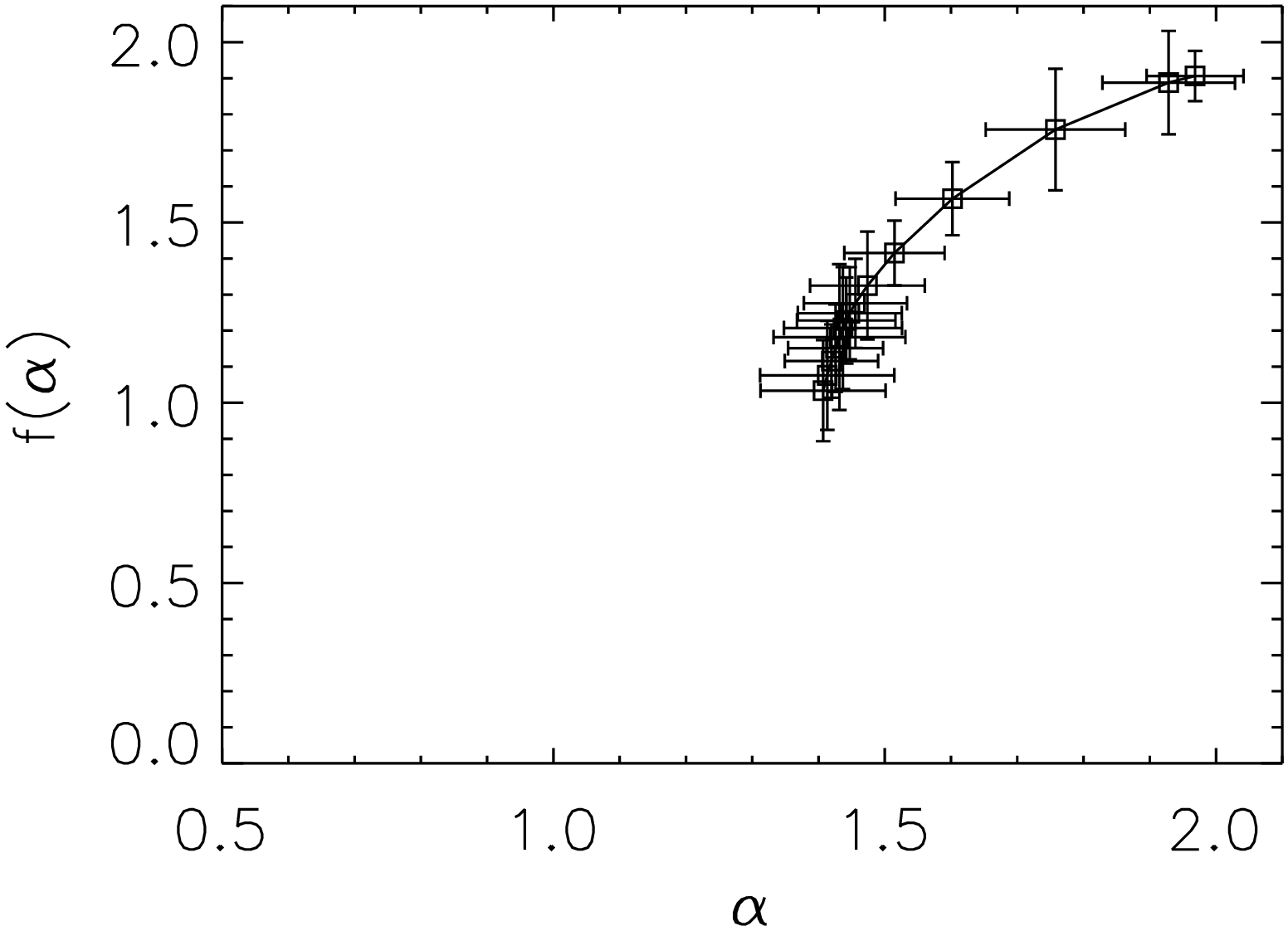}
\includegraphics[width=4.cm]{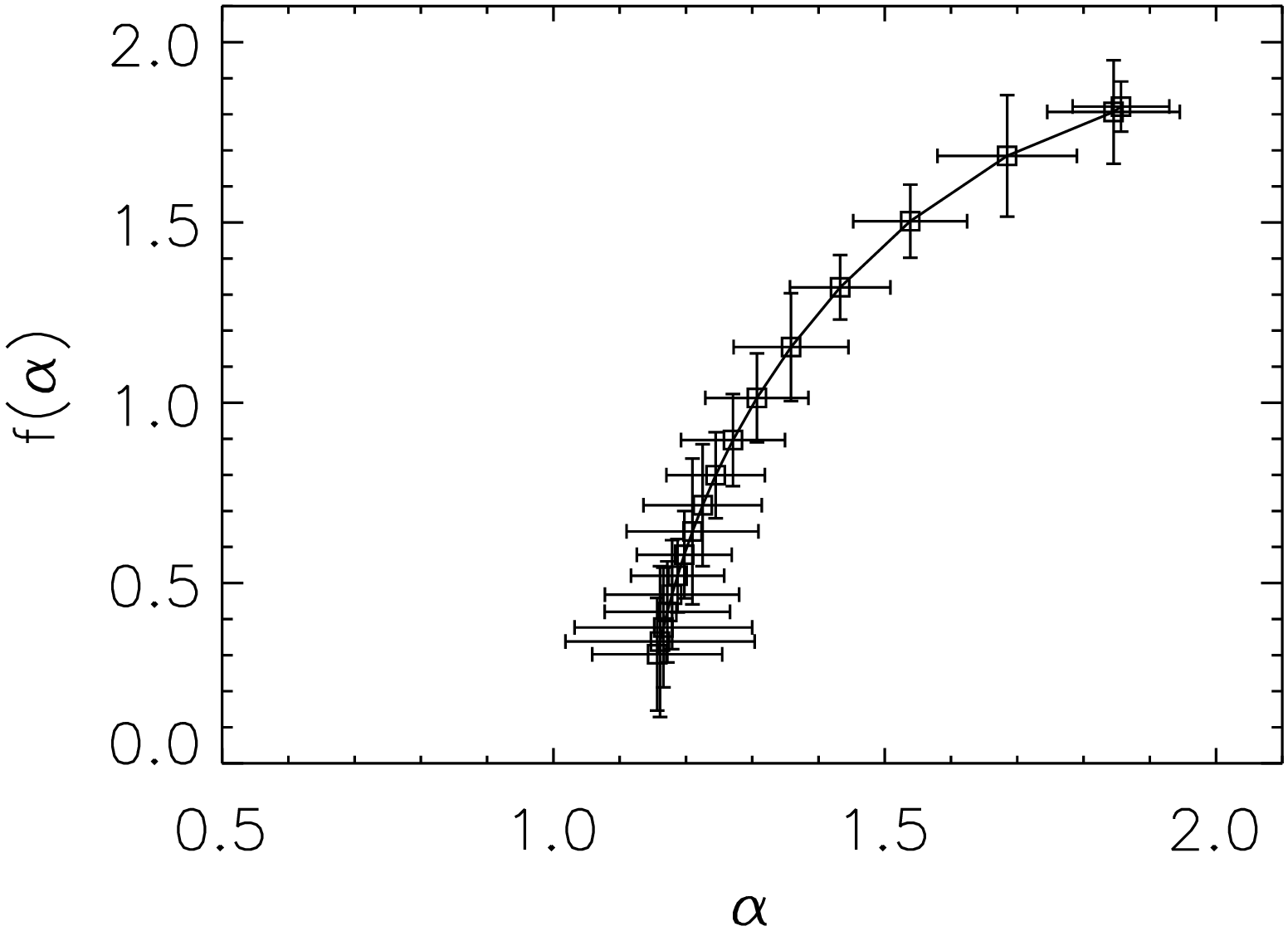}
\includegraphics[width=4.cm]{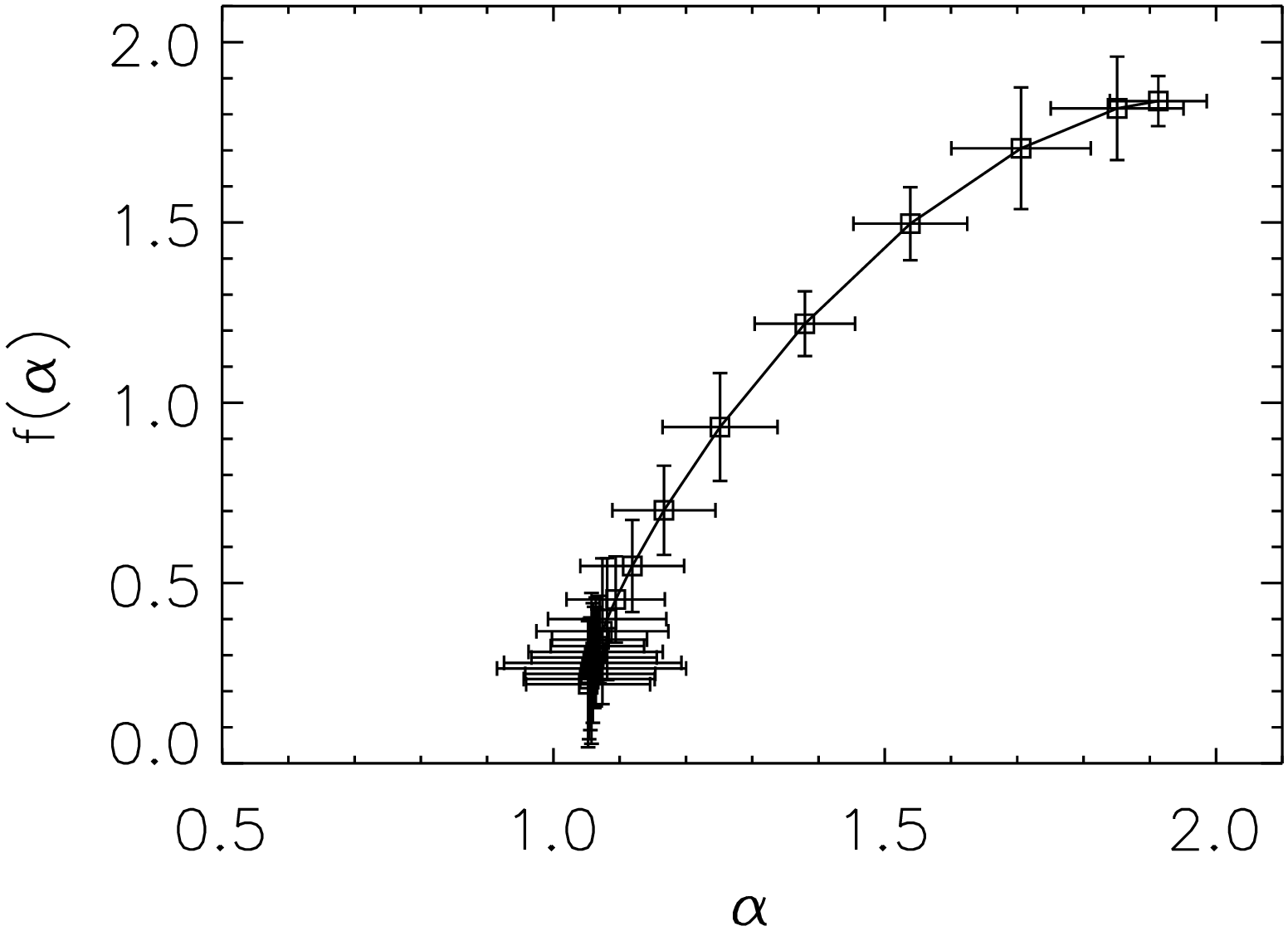}\\
\includegraphics[width=4.cm]{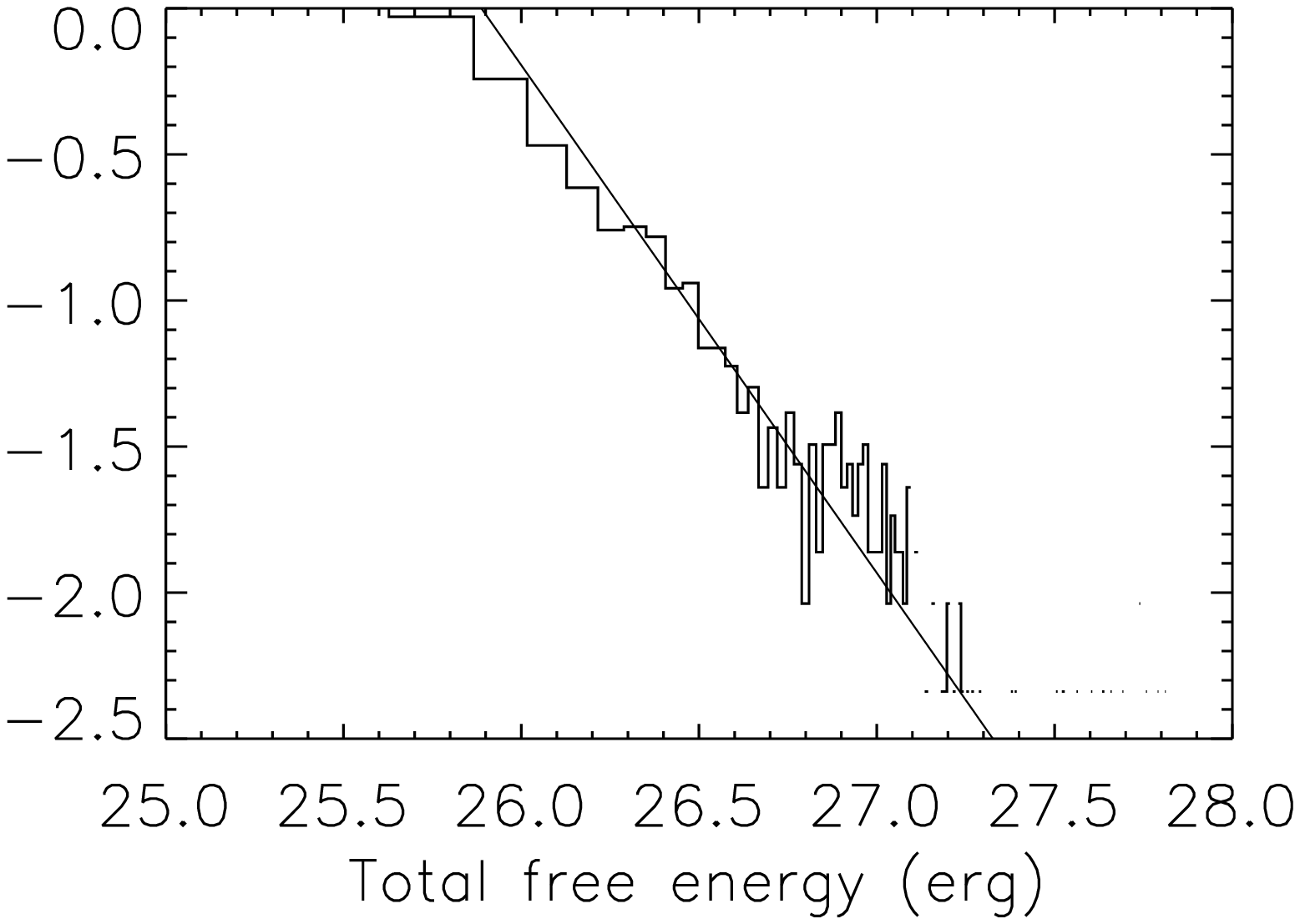}
\includegraphics[width=4.cm]{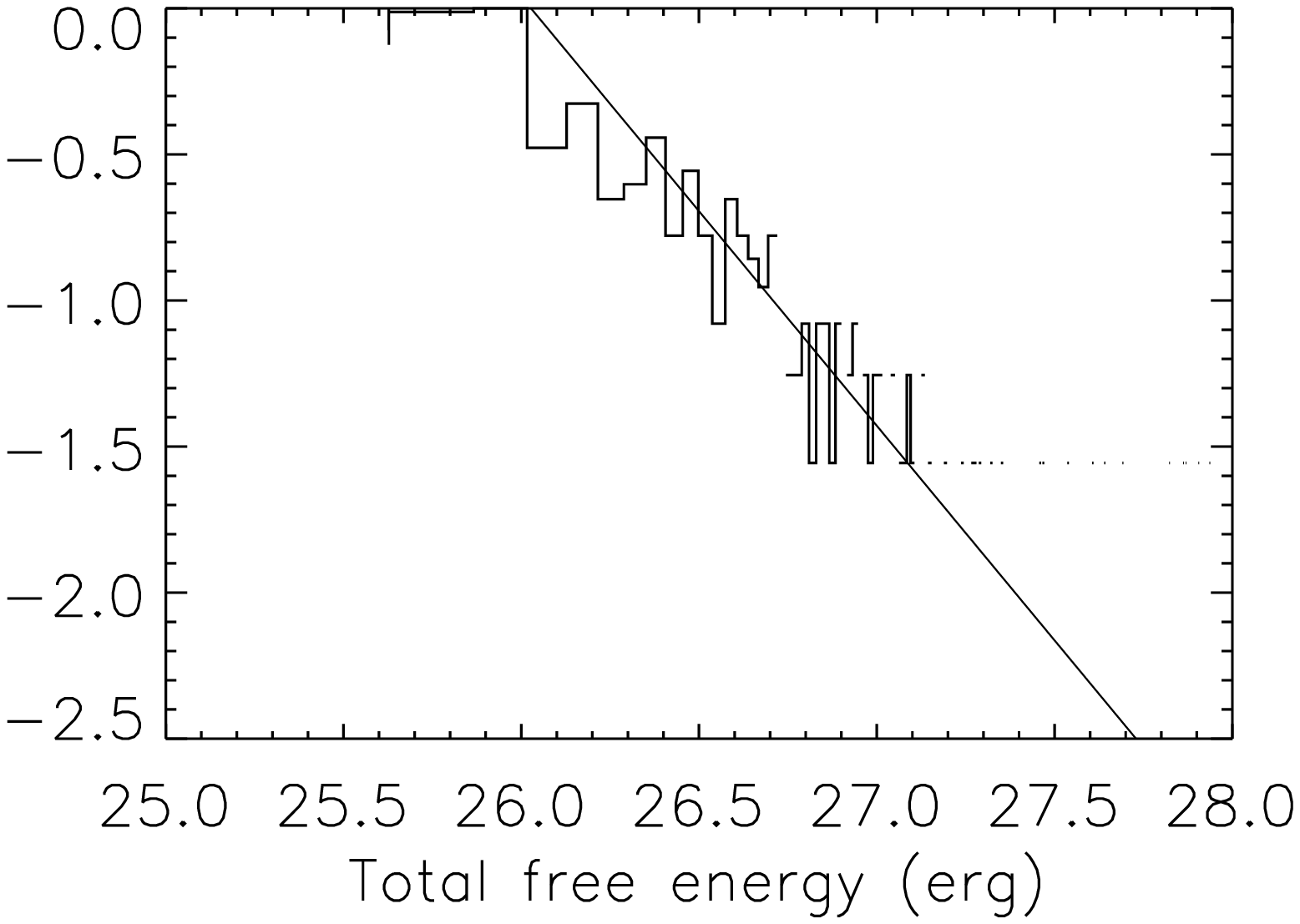}
\includegraphics[width=4.cm]{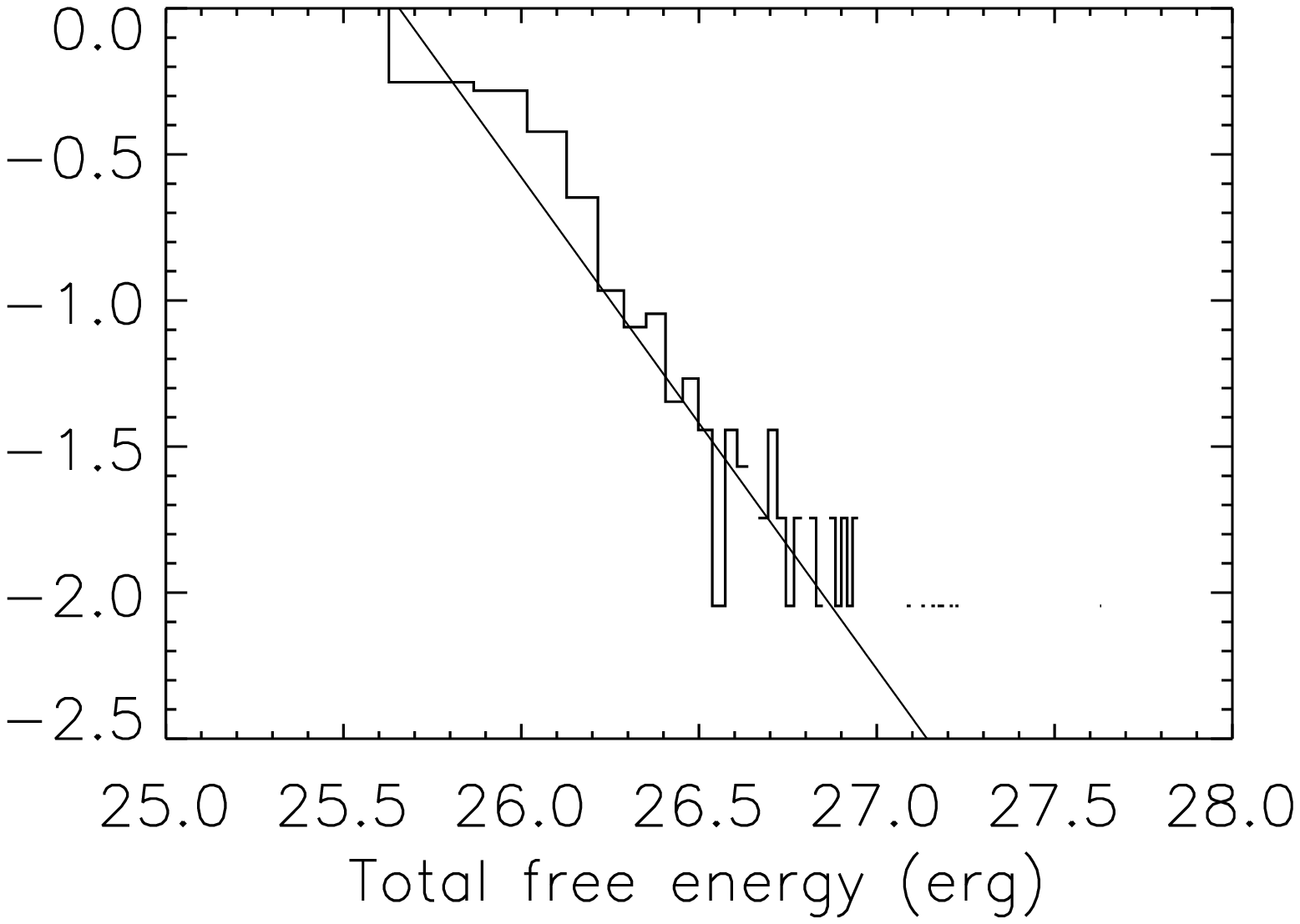}
} 
\caption{Top: Details of  MDI magnetograms, acquired on 05 June 2000 00:00 UT (left), 08 June 2000 04:48 UT (center), 10 June 2000 16:00 UT (right), showing the decay phase of NOAA 9026. Middle: corresponding multifractal spectra. Bottom: corresponding distribution functions of total free energy. }
\label{fig_9026}
\end{figure*}

\begin{figure*}
\centering{
\includegraphics[width=4.cm]{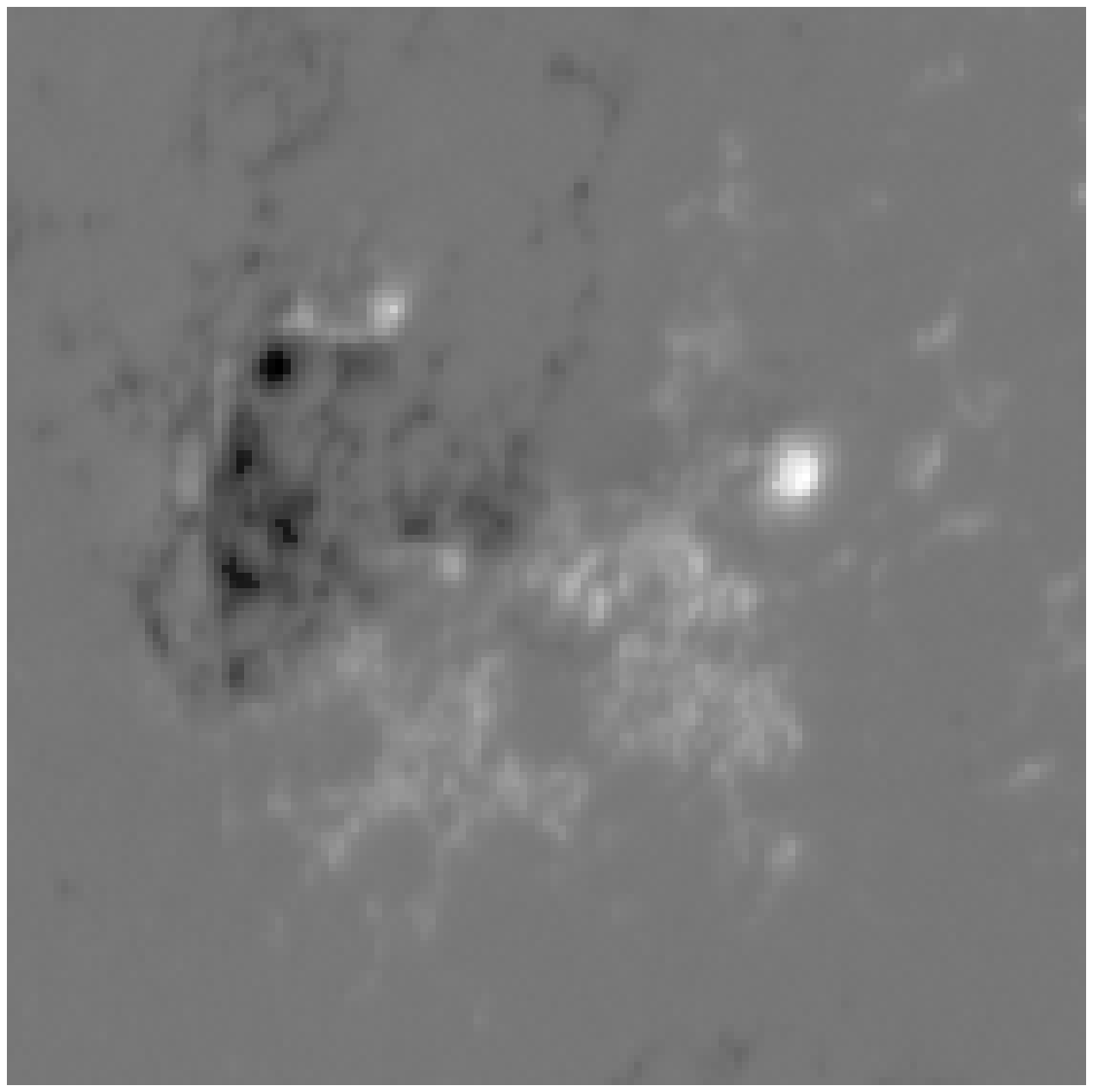}
\includegraphics[width=4.cm]{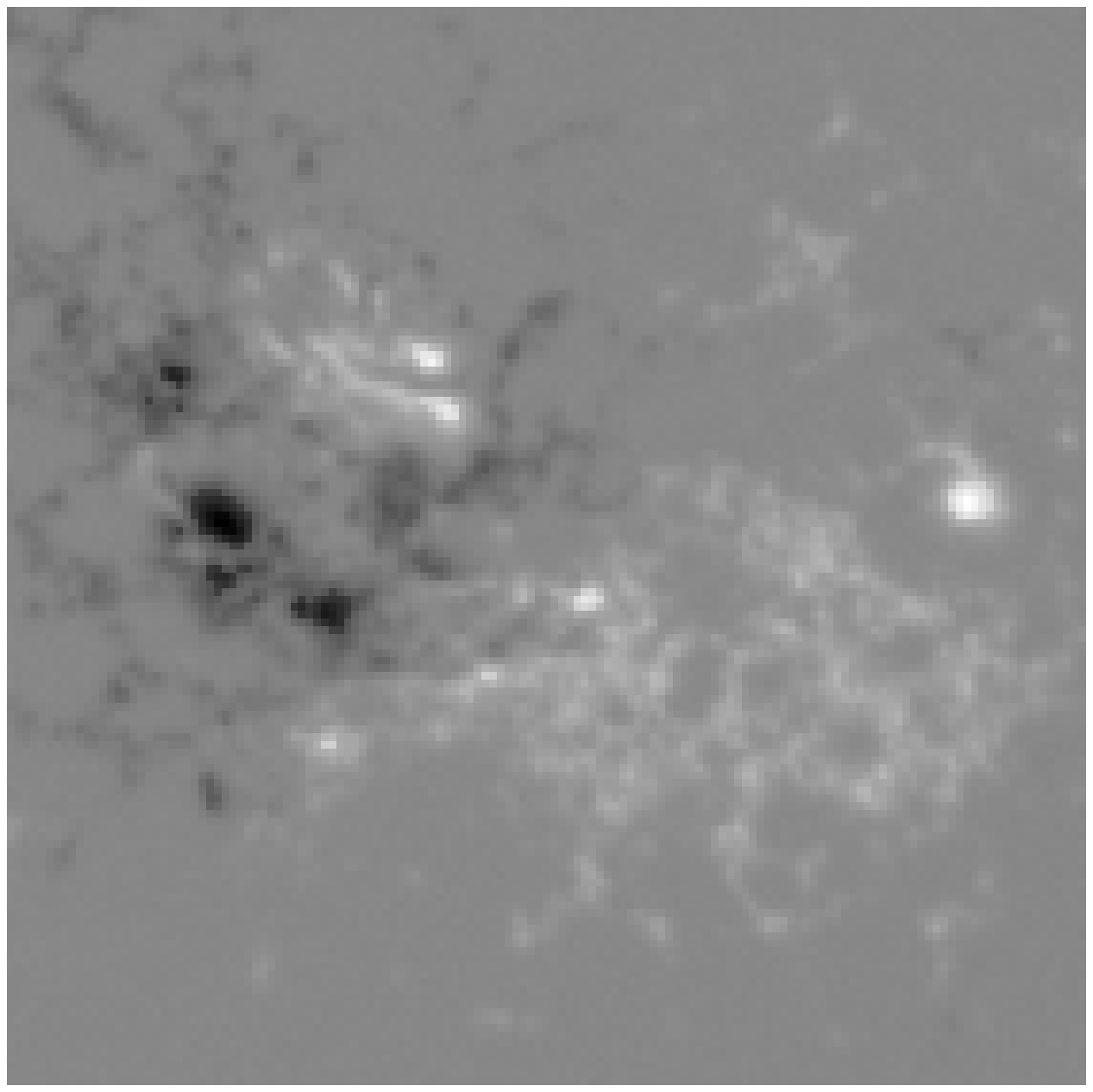}
\includegraphics[width=4.cm]{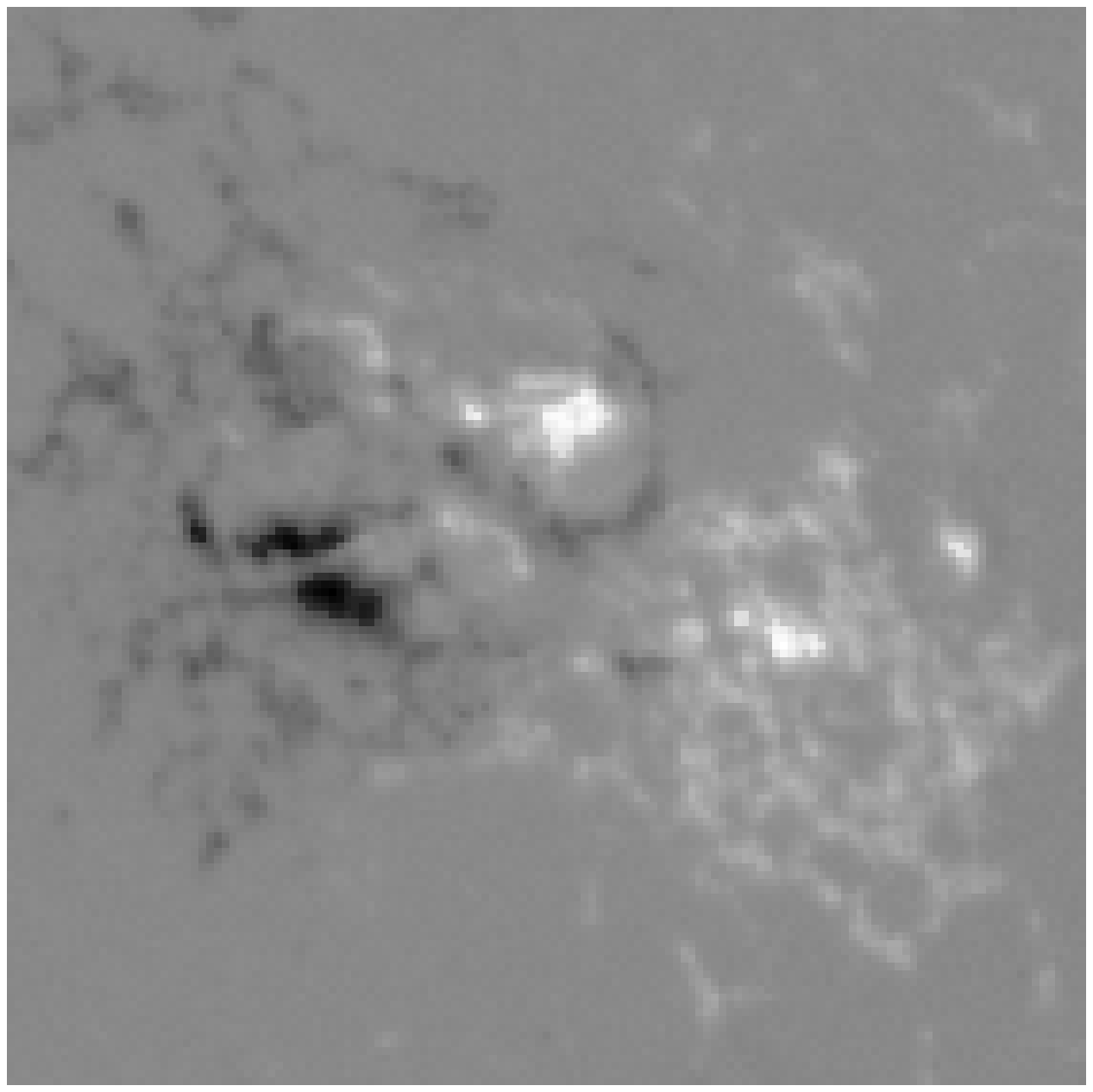}\\
\includegraphics[width=4.cm]{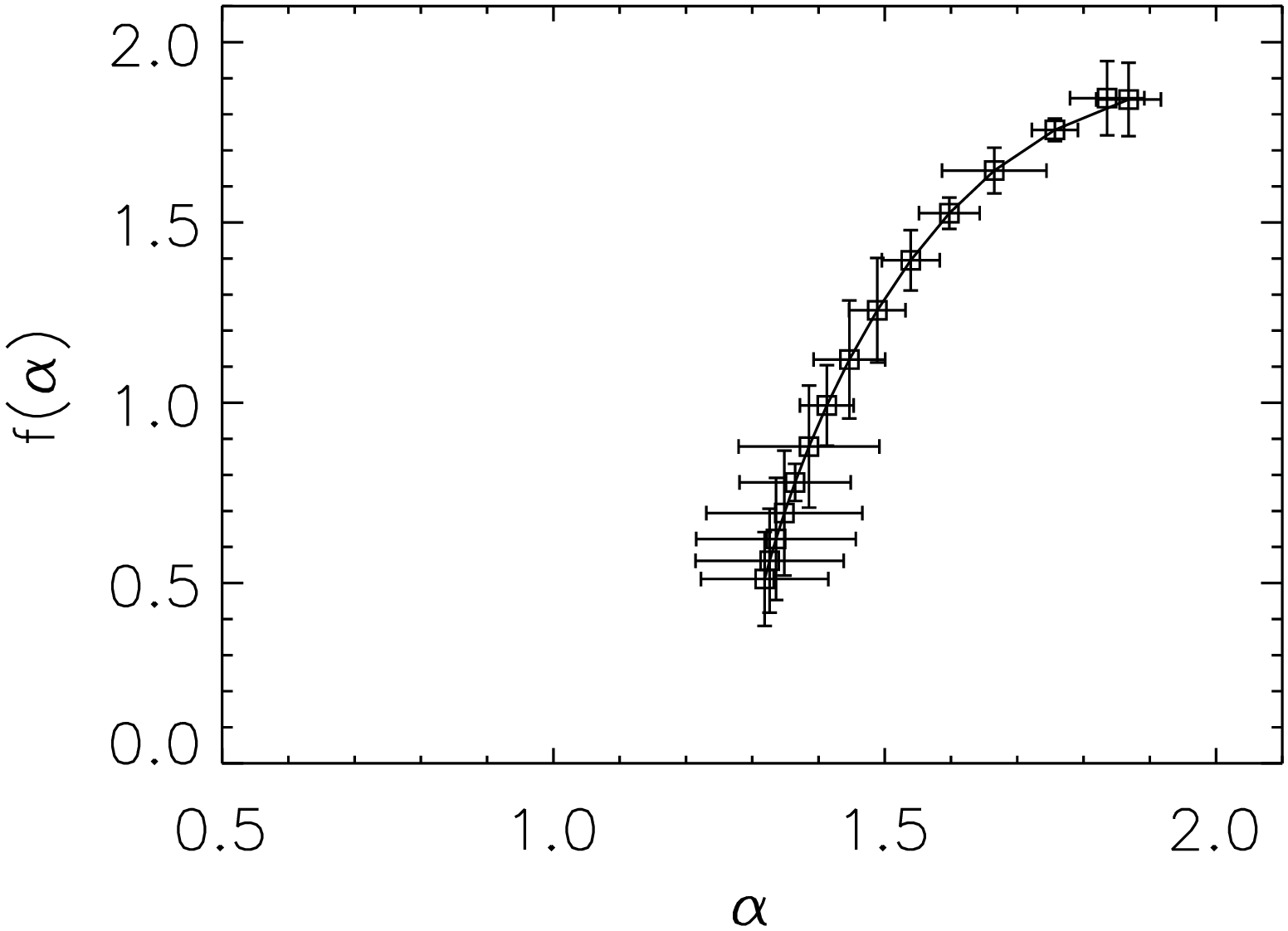}
\includegraphics[width=4.cm]{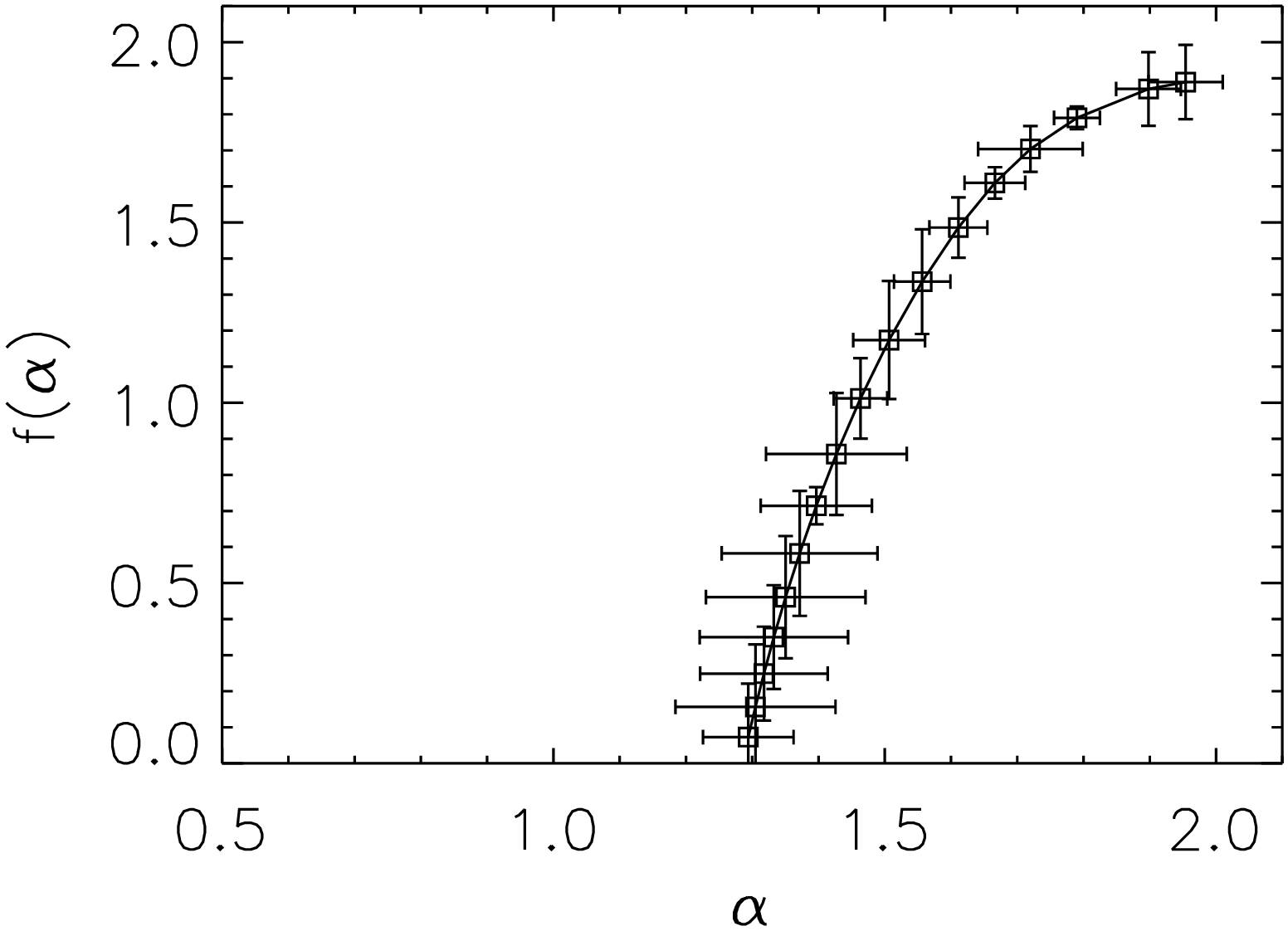}
\includegraphics[width=4.cm]{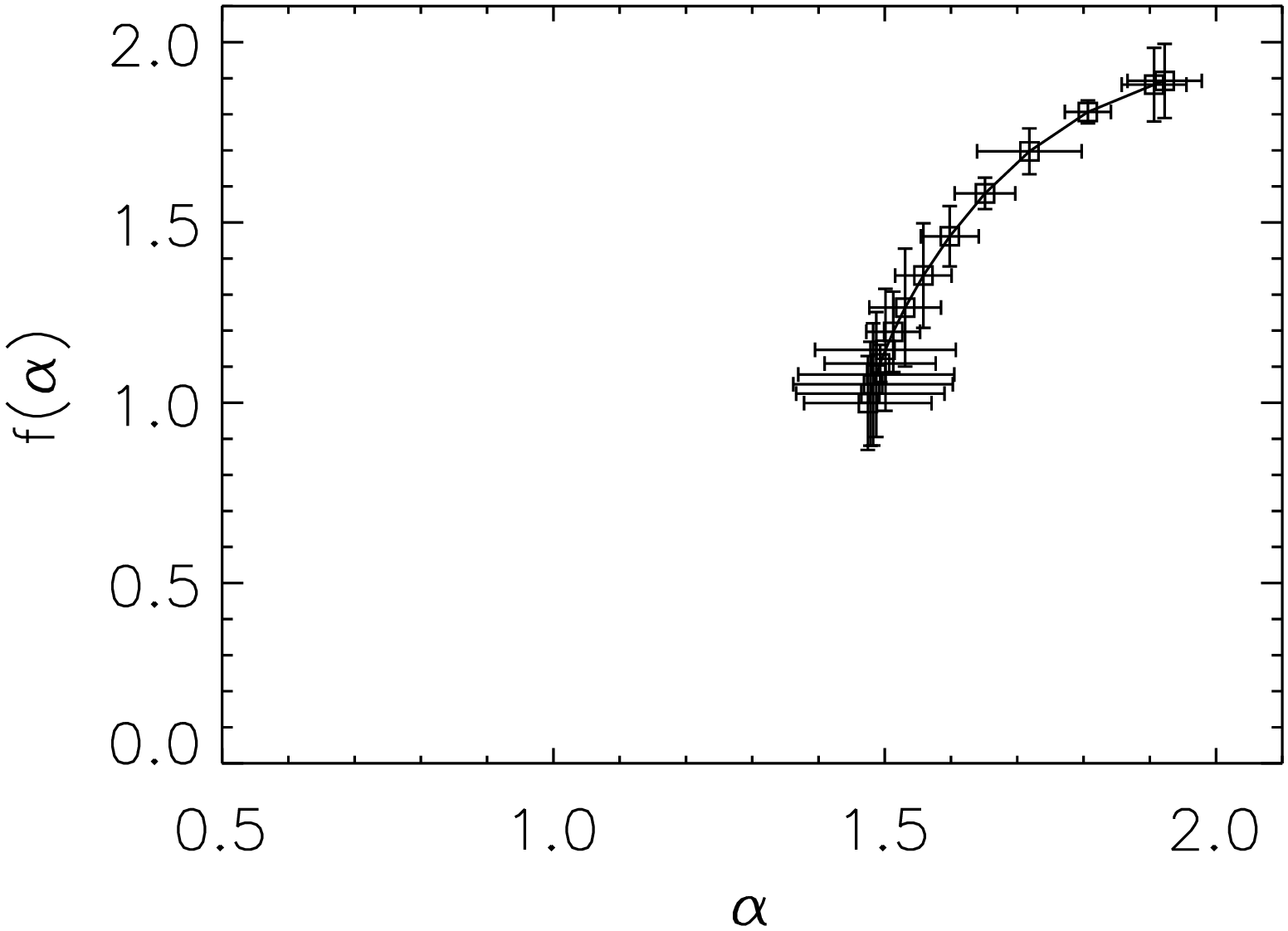}\\
\includegraphics[width=4.cm]{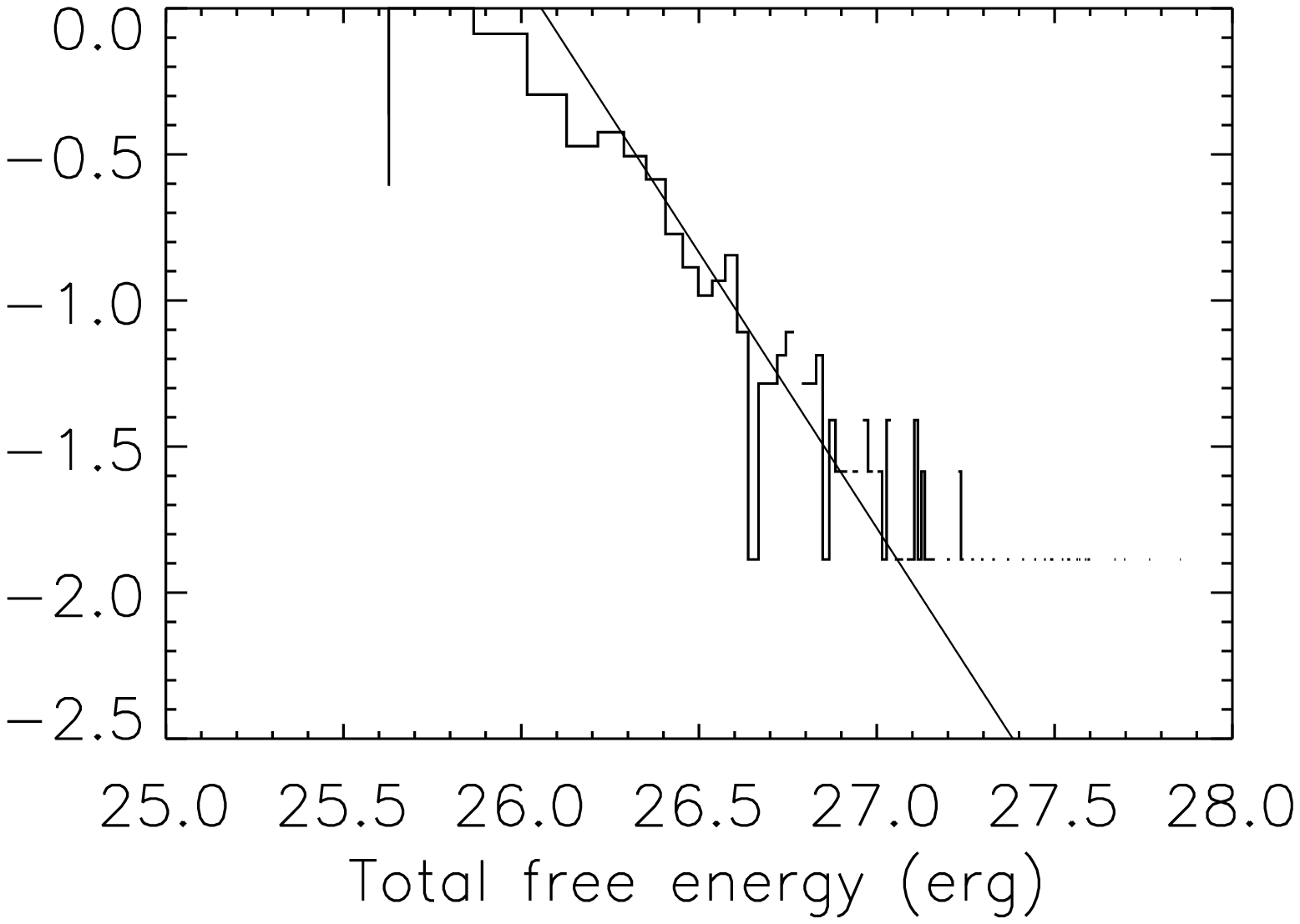}
\includegraphics[width=4.cm]{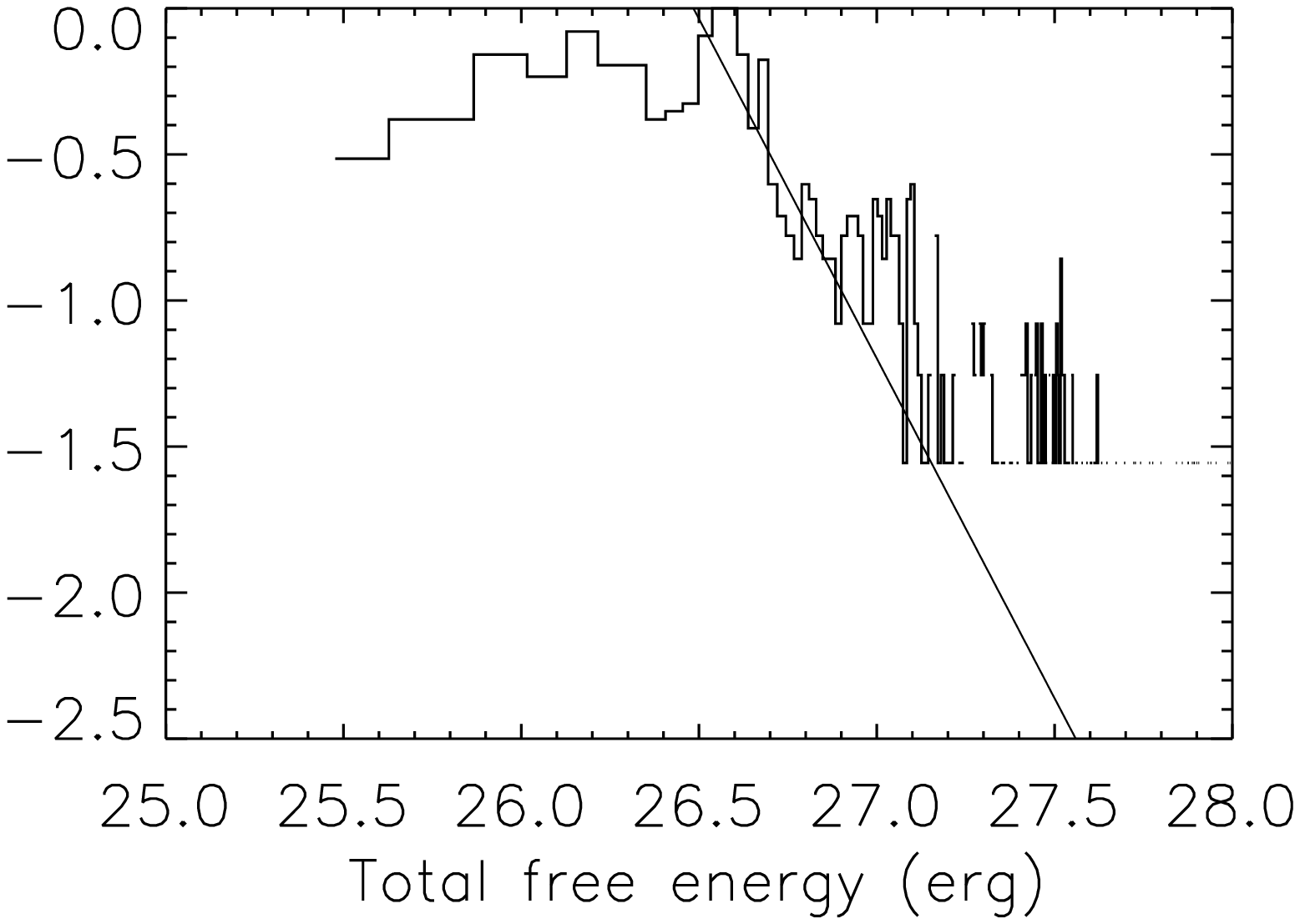}
\includegraphics[width=4.cm]{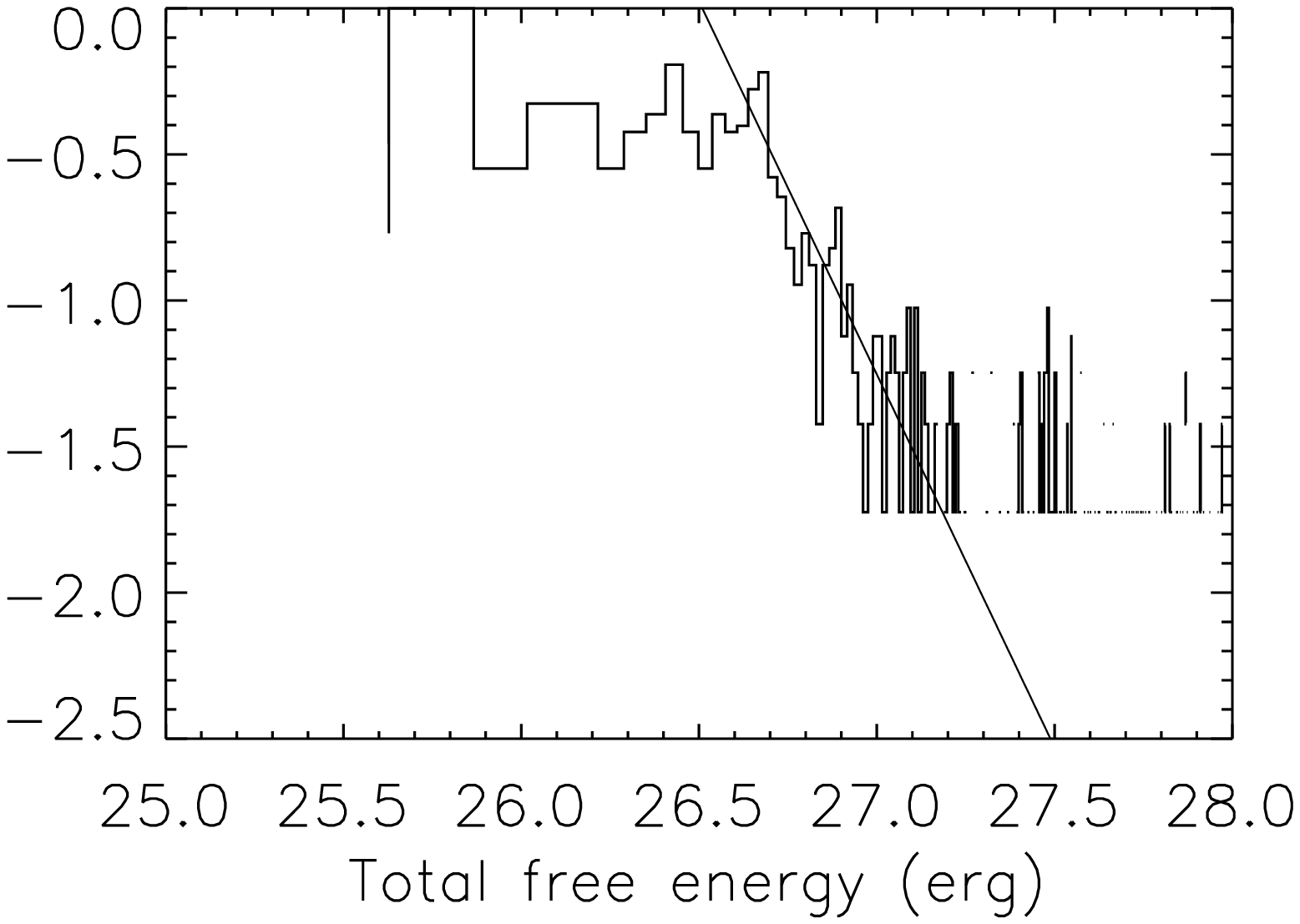}
} 
\caption{Top: Details of  MDI magnetograms, acquired on 11 December 2001 06:27 UT (left), 14 December 2001 09:36 UT (center), 16 December 2001 09:36 UT (right), showing the temporal evolution of NOAA 9733. Middle: corresponding multifractal spectra. Bottom: corresponding distribution functions of total free energy. }
\label{fig_9733}
\end{figure*}

Considering the distribution function of the total free energy $E_{i}$ in the selected discontinuities, we obtained a power law with a different power law index $\alpha_{E}$ for each magnetogram. This indicator was estimated by a linear fit to the logarithm of the total free energy distribution. Since the shapes of the distribution functions evolve with time (see for instance Figs. \ref{fig_9236}, \ref{fig_9026} and \ref{fig_9733}), we employed an automatic technique \citep{mcateer2005} to determine the linearity ranges and perform the linear fits. 

The difference between $I_v$ and  $\alpha_{E}$ estimated using force-free parameters obtained by the best fit of the observed loops and using the resonant values are on average $10\%$ for both quantities. These differences are lower than temporal variations measured for $I_v$ and are within the error associated with the estimation of $\alpha_{E}$ (see \S4.1), thus showing that the choice of using as a force-free parameter the magnetogram resonant value produces reasonable results. 
 
%------------------------------------------------------------------------------

%-----------------------------------------------------------------------
%-------------------------------------------------------------
%________________________

%-------------------------------------------------------------
%--------------------------------------------------------------------------

\section{Results}

\subsection{Temporal evolution of the indicators} 

We studied the temporal evolution of $D_{8}$, $C_{d}$, $D_{d}$, $I_{v}$ and $\alpha_{E}$ for each superactive region. For some indicators, we found these variations to be dependent on the evolution of the ARs. 
\begin{figure*}

\centering{
\includegraphics[width=8cm]{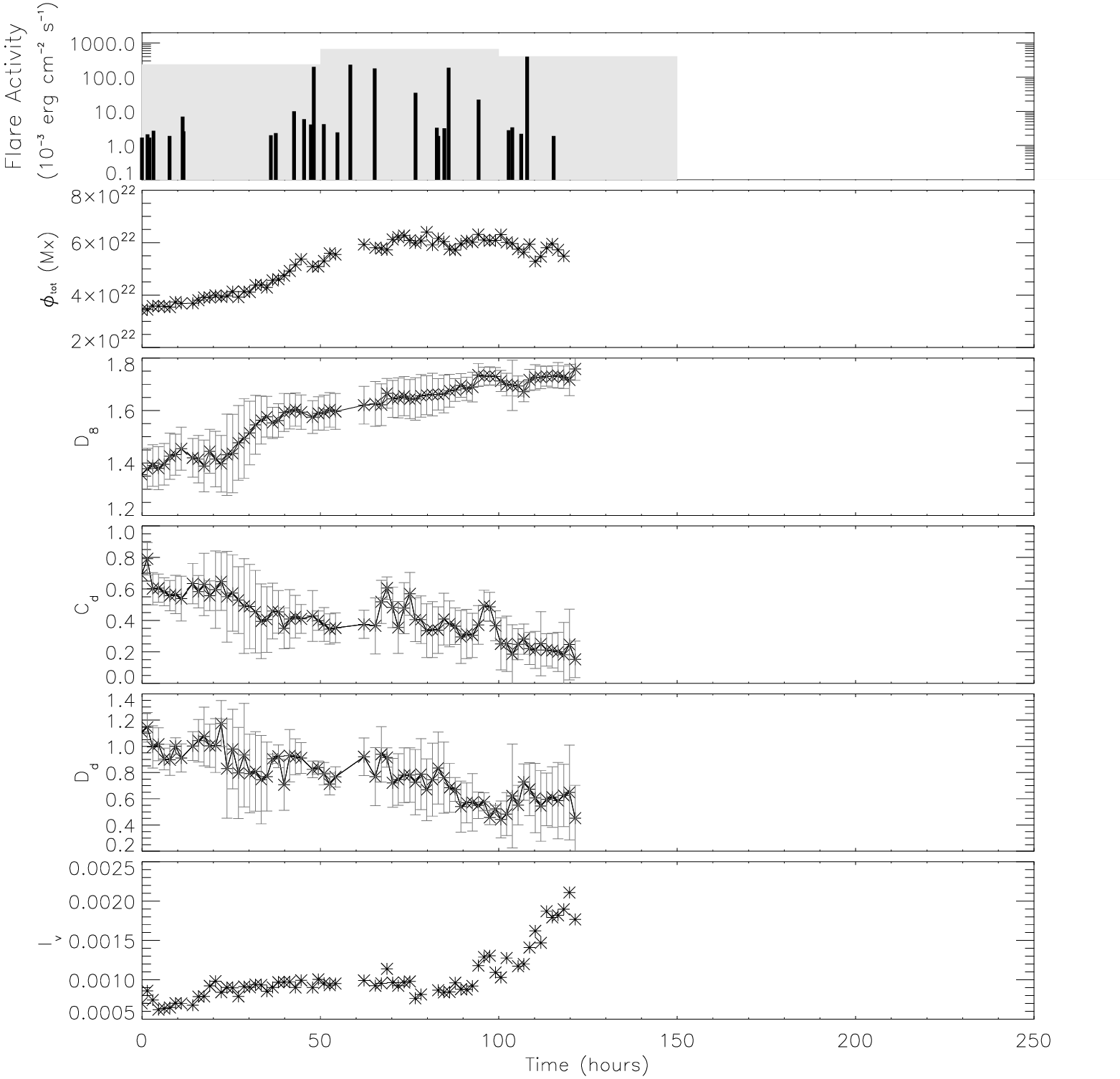}
\includegraphics[width=8cm]{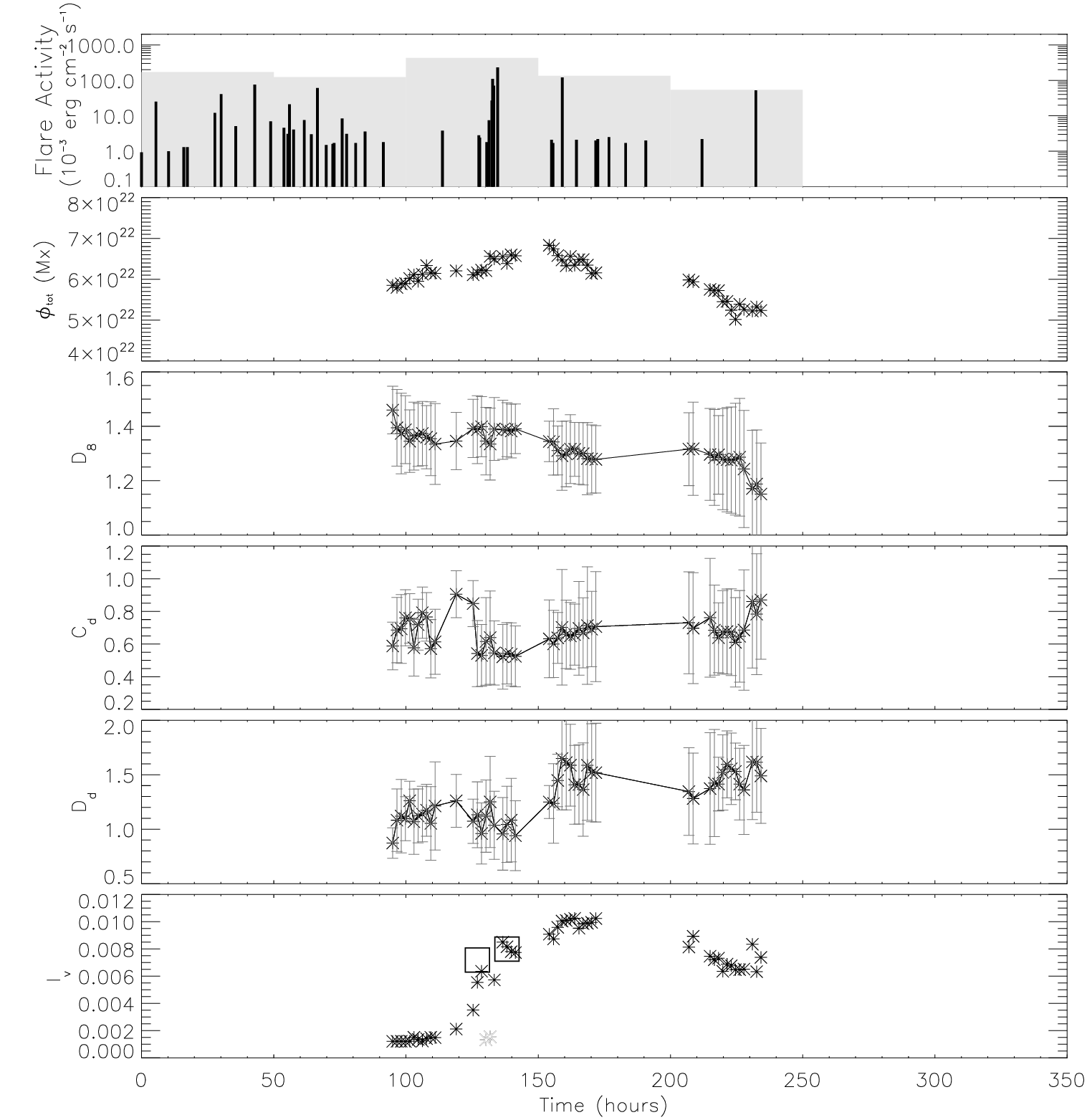}
\hspace*{0.1cm}{\bf(a)}\hspace*{7.2cm}{\bf(b)}

} 
\caption{ Temporal variations of the estimated parameters for (a) region NOAA 9236 and (b) NOAA 9026 (b). The squares in the last panel of (b) indicate the $I_{v}$ values deduced from the force-free field extrapolations performed using a best fit criterion with the loops observed by TRACE at that time.}
\label{fig_9236_index_var}
\end{figure*}

We found that for all the active regions characterized by an emerging magnetic flux, the multifractal spectrum 'contracts' and gradually shifts toward higher values of both $\alpha_q$ and $F(\alpha_q)$. As shown for instance in Fig. \ref{fig_9236} for NOAA 9236, the MDI magnetograms taken on 22, 23 and 26 November 2000 (first row) show the emergence of flux in photosphere, and the corresponding plots of multifractal spectra  (second row) show that the values of both $F(\alpha_q)$ and $\alpha_q$ increase, while the ranges of values spanned by these quantities gradually reduce. As already pointed out by \citet{conlon2008}, who found similar results, this reflects the fact that magnetic flux typically emerges fragmented in small-size features, which coalesce and evolve into larger and more complex structures. 
%The opposite is found during the decay phase, NOAA 9026. As shown in Fig. %\ref{fig_9026}, in fact, for NOAA 9026 both $\alpha_q$ and $F(\alpha_q)$ %decrease, while the spectrum broadens. 
On the contrary, during the decay phase both $\alpha_q$ and $F(\alpha_q)$ decrease while the spectrum broadens, as illustrated in Fig. \ref{fig_9026} for NOAA 9026. 
This probably reflects the fragmentation of the magnetic field into less and less complex small-scale regions as confirmed by the magnetograms shown in the first raw of Fig. \ref{fig_9026}.

For ARs in the emerging phase the fractal dimension $D_{8}$ increases with time, while both multifractal indices $D_{d}$ and $C_{d}$ decrease, as shown for instance in Fig.\ref{fig_9236_index_var} (a). The opposite is found for NOAA 9026 during its decay phase, as shown in Fig.\ref{fig_9236_index_var} (b). In these plots t=0 corresponds to the time when the first flare was registered.

For regions observed around the maximum of their life, the evolution of the multifractal spectrum is more complex. In general, we found that the fractal dimension stays fairly constant, while larger variations of multifractal indicators are observed at the emergence of new magnetic flux. This is particularly evident, for instance, for NOAA 10030, as shown in the appendix. Some cases also show that at the very beginning of the decay phase, that is, during the hours after the maximum of flux is measured, the fractal dimension continues to increase and multifractal indicators to decrease. Figure \ref{fig_9733} displays for instance the magnetograms  and the corresponding multifractal spectra of NOAA 9733 at the beginning of the observations (left), at the moment in which the maximum total flux was measured (center), and at the end of the observations (right). 
Comparison of images in the left and middle columns shows that with the emergence of new magnetic flux, both the number of compact larger features and of fragmented smaller features increases. The former result from the coalescence of 'old' flux into larger and compact features, whereas the latter are the signature of 'new' emerged flux. As a consequence, the multifractal spectrum broadens and $f(\alpha_{min})$, i.e. the fractal dimension of the largest scale structures, decreases. 
The right column shows that, at the very beginning of the decay phase, the  small features coalesce into larger and more complex regions, so that the spectrum 'contracts' and shifts toward higher values of both $f(\alpha_q)$ and $\alpha_q$, although the total amount of magnetic flux is decreasing.
 
On the other hand, the temporal evolution of $I_v$ does not seem to be strictly correlated with the evolution of the magnetic flux in the photosphere. Although in some cases during the emerging phases $I_v$ increases, as shown in Fig.\ref{fig_9236_index_var} (a), and during the decaying phases it decreases, as shown in Fig.\ref{fig_9236_index_var}(b), in other cases the amount of magnetic discontinuities changes independently from the magnetic flux observed in the photosphere. This result confirms that the coronal field extrapolated using the linear force-free approximation provides some independent information from the magnetic field above the level of the measurements. We also found that for most of the regions analyzed, $I_v$ does not vary monotonically, but it is rather a discontinuous function of time. 
 Square symbols in the plot of $I_v$ in Fig.\ref{fig_9236_index_var} (b) represent the total unstable volume computed by using a force-free parameter derived from the comparison with TRACE data. The values obtained in this way are very close to the ones obtained when the force-free field parameter is fixed just below its resonant value. 

The analysis of the temporal evolution of the distribution functions of total free energy of unstable volumes shows that during the emergence phase the peak of the distribution, as well as the linearity range, shift gradually toward higher energy values, as shown for instance in Fig.\ref{fig_9236} and Fig.\ref{fig_9733}. On the contrary, temporal variations of the energy distribution of region NOAA 9026 (Fig.\ref{fig_9026}), observed during the decay phase, do not seem to have a particular trend with the evolution of the region. Variations with time of the slope of the linear fits to the distributions ($\alpha_E$ indicator) are within the error bars and are not correlated with the evolution of the regions.

\subsection{Correlation with the flare index}

In order to investigate the correlation of the magnetic field topology with the flare occurrence we reported the flare index and the other parameters in the same plots for each superactive region (see for instance Figs. \ref{fig_9236_index_var} (a) and (b) and Appendix A).  
We represented each single flare with a black bar whose height is proportional to the X-ray flare peak observed by GOES and the total of the flare index over 50 hours with grey bars. We used the logarithmic scale to show the magnitude of different flares in the same plot.

We estimated for each AR the correlation coefficients between the considered indicators and the flare index computed over windows of 50 h. Results are reported in Table \ref{tab1} (columns from 5 to 8).

In approximately 50\% of the ARs we found that $D_{8}$ is positively correlated with the flare index: 7 ARs have a correlation coefficient larger than 0.5, with the highest value of 0.88 for NOAA 10656. For the same ARs (except for NOAA 10375) we obtained that $D_{d}$ and $C_{d}$ are anticorrelated with the flare index. The smallest correlation coefficients between the multifractal parameters and the flare index are -0.78 and -0.66 for NOAA 10365. For other regions the correlation between fractal and multifractal indicators is almost null. A comparison to total flux variation (see Appendix A) reveals that most of these regions have been observed at about the maximum of their evolution. The peculiar values found for NOAA 9733 are due to the large decrease of the flare index at the beginning of the decay phase, during which, as already pointed out in previous paragraph, $D_{8}$ increases while $D_{d}$ and $C_{d}$ decrease.

\begin{figure}
\centering{
\includegraphics[width=7.7cm]{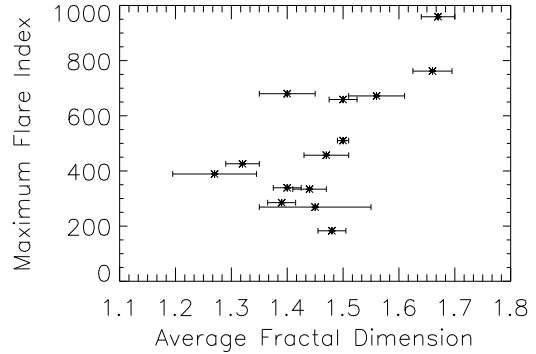}} 
\caption{Maximum of the flare index versus average generalized fractal dimension $D_{8}$ for the 14 ARs analyzed.}
\label{fig_FlInd_vs_Df}
\end{figure}

Figure \ref{fig_FlInd_vs_Df} displayes the relation between the average fractal dimension $D_{8}$ and the maximum measured for the flare index. In agreement with previous works \citep{mcateer2005,abramenko2005b}, the plot suggests that more complex regions have a higher probability of developing more intense and a larger number of flares. Nevertheless, the large scatter of data does not allow us to derive an analytical relation between $D_8$ and the flare index. Note that the correlation coefficient between these two quantities is 0.7, while the correlation coefficients between the maximum flare index and the average values of $D_{d}$ and $C_{d}$ are approximately -0.7. Interestingly, the correlation is 0.5 if $D_{8}$ is averaged over the same 50 h time interval. 

From Table \ref{tab1} we can see that there is a large scatter of correlation coefficients between flare indices computed over windows of 50 hours and $I_v$. On the other hand, we noted that $I_v$ shows more rapid variations than the fractal and multifractal parameters. Clear and strong increases of its value correspond to phases of particularly frequent and intense activity, as we can see around 100 hours and 120 hours from the beginning of our observation in NOAA 9236 and NOAA 9026, respectively (Fig. \ref{fig_9236_index_var} (a) and (b))

As reported also by \citet{vlahos2004} $\alpha_E$ does not show a particular trend with time, but it is interesting to notice that the average value estimated during the observation time for our superactive regions, except for NOAA 9733 and NOAA 10486, is between -1.6 and -1.8 (see Table \ref{tab1}, last column).

%-------------------------------------------------------------
%____________________________________________________________ 

\section{Discussion and conclusions}

We investigated the temporal evolution of the generalized fractal dimension, the contribution diversity, the dimensional diversity, the volume occupied by the potentially unstable regions and the power law index of the distribution function of the total free energy in 14 superactive regions. The first three quantities provide information about complexity of the magnetic field in the photosphere, while the other two provide information about magnetic field organization in the upper layers of the solar atmosphere. In particular, we investigated the variations of these indicators with the emergence of magnetic flux. With the exception of the  fractal dimension, we have also employed these indicators for the first time  for statistical studies aimed at the investigation of their correlation with flare productivity.

We found that fractal dimension and multifractal parameters are anticorrelated. During the emergence phase, the fractal dimension rapidly increases and multifractal parameters decrease. The opposite was observed during the decay phase. In the 'mature' phase, fractal dimension stays fairly constant, while larger variations have been observed for multifractal parameters. Movies reveal that these variations are usually due to the emergence of new magnetic flux. These findings are in agreement with results presented by \citet{conlon2008}, who also observed a rapid increase of multifractal parameters at the very beginning of the emergence phase. We do not find such an increase in our results, since we analyzed images taken after the observation of the first flare, that is, after the emergence of a large part of the magnetic flux. These results confirm the fact that magnetic flux emerges fragmented in the photosphere and then it evolves in more and more compact and complex regions. 
Moreover, the evolution of the magnetic field in the photosphere seems to influence the volume occupied by the magnetic discontinuities, that is the value of $I_v$ and the distribution functions of the energy of unstable volumes, only in part. Of course, the more flux that emerges, the greater is the probability of formation of unstable regions. Nevertheless, the complexity of the magnetic field in the upper layers of the solar atmosphere can be largely increased by photospheric magnetic field dynamics \citep{Choe2000}, that fractal and multifractal studies do not properly take into account for. 

We also analyzed the temporal evolution of the distribution functions of the total energy of unstable volumes. We found that for most of the observed regions, during the emergence phase the distribution function peaks as well as the linearity range shift toward higher energy values, while the slope of the linear fit (parameter $\alpha_E$) does not vary significantly with the evolution of the region.

In 6 ARs we found that the flare index is positively correlated with $D_8$ and anticorrelated with  $C_d$ and $D_d$. This indicates that in many cases the photospheric complexity is a strong condition for flare production. On the other hand, the absence of any correlation in the other ARs confirms that the evolution of the photospheric magnetic field investigated by the fractal and multifractal parameters might not clearly reflect the real complexity of the AR in the upper levels of the solar atmosphere. 

In agreement with previous works of \citet{mcateer2005} and \citet{abramenko2005b}, we also found that more complex regions are more likely to develop flares. In particular, \citet{abramenko2005b} found that the power spectrum index of ARs magnetograms and the daily flare index are correlated and that their variation can be described by a power law. We do not find such a law in our data (see Fig.\ref{fig_FlInd_vs_Df}), most likely because our sample includes only superactive regions. Nevertheless, we also find good correlation between flare index and average fractal dimension of a region, and good anticorrelation between flare index and average multifractal indicators.    

For $I_v$ we did not find a significant correlation with the flare index computed over time windows of 50 hours, but we noted a clear increase of this parameter during the phases when the ARs showed a burst of flares. This suggests that the presence of magnetic discontinuities plays an important role in the determination of the conditions useful for the occurrence of magnetic reconnection. Moreover, the fact that the flares occur more frequently while the volume of potentially unstable regions increases, indicates that the flare itself does not provide any mechanism to stop the formation of the instability regions. Therefore, we believe that the source of the current sheet formation should not be searched for only inside the discontinuities, but in the whole AR evolution.

Although we did not observe any significant variation with time of $\alpha_E$, we obtained average values of this indicator below the values obtained by \citet{vlahos2004} over two quiet ARs. 
This confirms the hypothesis of these authors according to which the steeper the slope of the free energy distribution function of unstable volumes of an AR, the higher is the probability of flare production.

We conclude that among the complexity parameters analyzed and for the time resolution of the dataset, the generalized fractal dimension $D_d$ and the potential unstable volume of magnetic discontinuities $I_v$ are the most suitable measurements to investigate the flare activity over longer (50 hours) and shorter (few hours) time intervals, respectively. Moreover, $C_d$ is less useful and is redundant if $D_d$ is already estimated. This result is in contrast with results obtained in previous studies, such as \citet{georgoulis2005, abramenko2005,abramenko2008} which reported variations in the degree of multifractality of ARs with flare productivity. These discrepancies are likely due to the different spatial and temporal resolution of our dataset and to the employment of different multifractal estimators. In particular, 
\citet{georgoulis2005} compared results obtained with different estimators and found that the structure functions and their inertial-scale scaling exponents show larger temporal variations than the spectrum of the generalized dimension, that is, the variation of $D_q$ (defined in eqn.\ref{fracdim}) with $q$, with flare occurrence. Since the multifractal spectrum is a function of $D_q$, our results confirm, through an analysis of a larger sample of data, results presented in \citet{georgoulis2005}. The different spatial and temporal resolution of the data employed in this study can also contribute to explain the discrepancy of our results. 

A better spatial and temporal resolution is needed to investigate in detail our findings. Therefore, the analyses of MDI high resolution or HINODE/SOT data could provide a good test for our results. Larger samples of data and statistical techniques more sophisticated than the correlation, such as the one proposed by \citet{barnes2008}, should be employed to evaluate the ability of complexity indicators to forecast flares.

\acknowledgements
This work was supported by the European Commission through the SOLAIRE Network (MTRN-CT-2006-035484), by the Istituto Nazionale di Astrofisica (INAF), by the Agenzia Spaziale Italiana (contract ASI I/01510710, Studio Esplorazione
Sistema Solare, sub-task 2210 Origine delle perturbazioni solari), by the Universit\`{a} degli Studi di Catania and by the Regione Lazio (progetto CVS). S.C also thanks I. Ermolli for useful discussions and support.

\appendix
\section{Parameter evolution}
 
\begin{figure*}[!h]
   \centering
   \includegraphics[height=14cm]{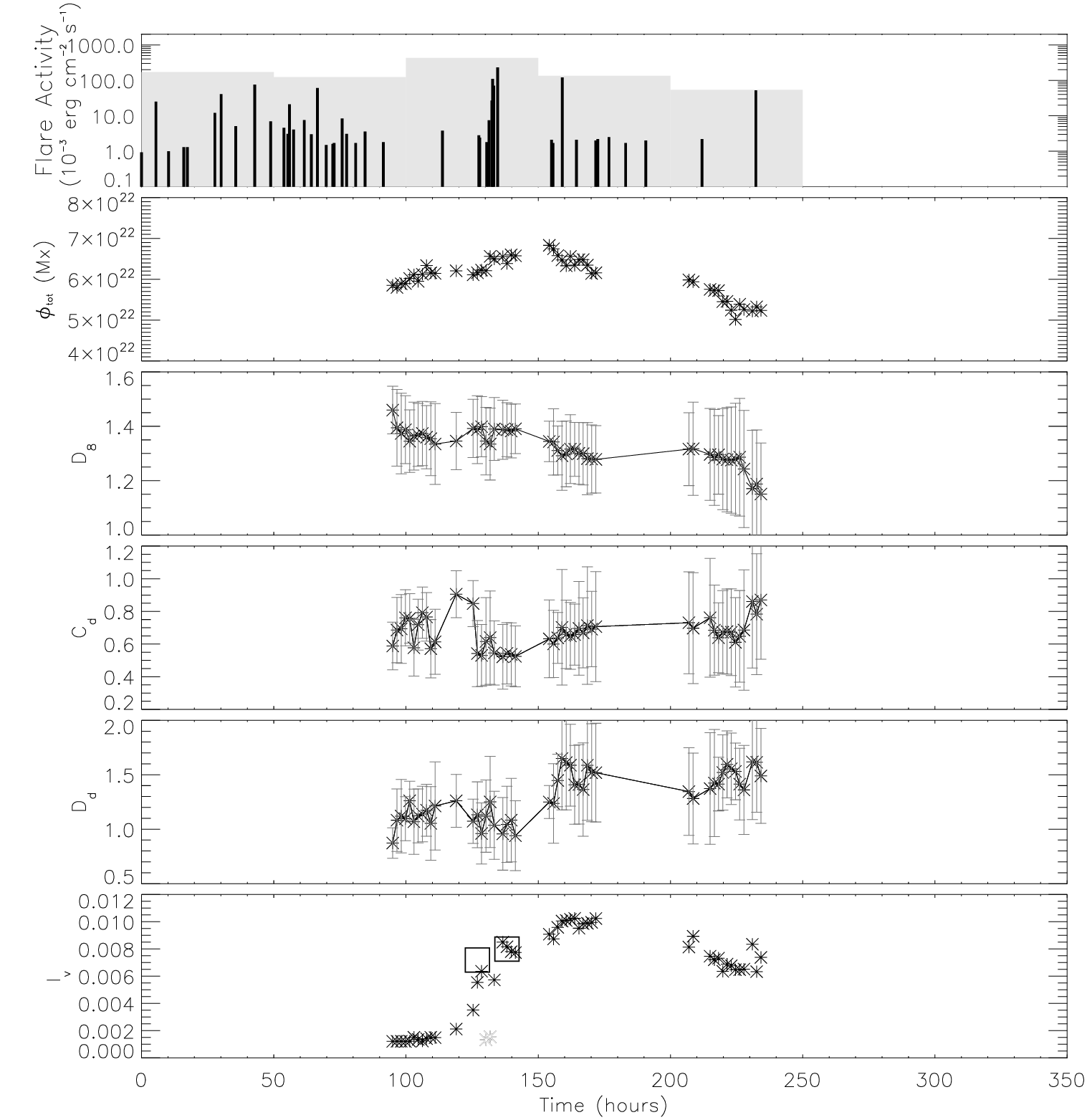}
   \caption{Flare activity, $\Phi_{tot}$, $D_8$, $C_d$, $D_d$ and $I_v$  vs. time for NOAA 9026}
   \label{Fig5}
\end{figure*}

In Figs. \ref{Fig5}-\ref{Fig18} we report the evolution of the flare activity, total magnetic flux $\Phi_{tot}$, generalized fractal dimension $D_8$, contribution diversity $C_d$, dimensional diversity $D_d$ and volume occupied by the potentially unstable regions $I_v$ for each of the 14 superactive regions.

\begin{figure*}[!h]
   \centering
   \includegraphics[height=14cm]{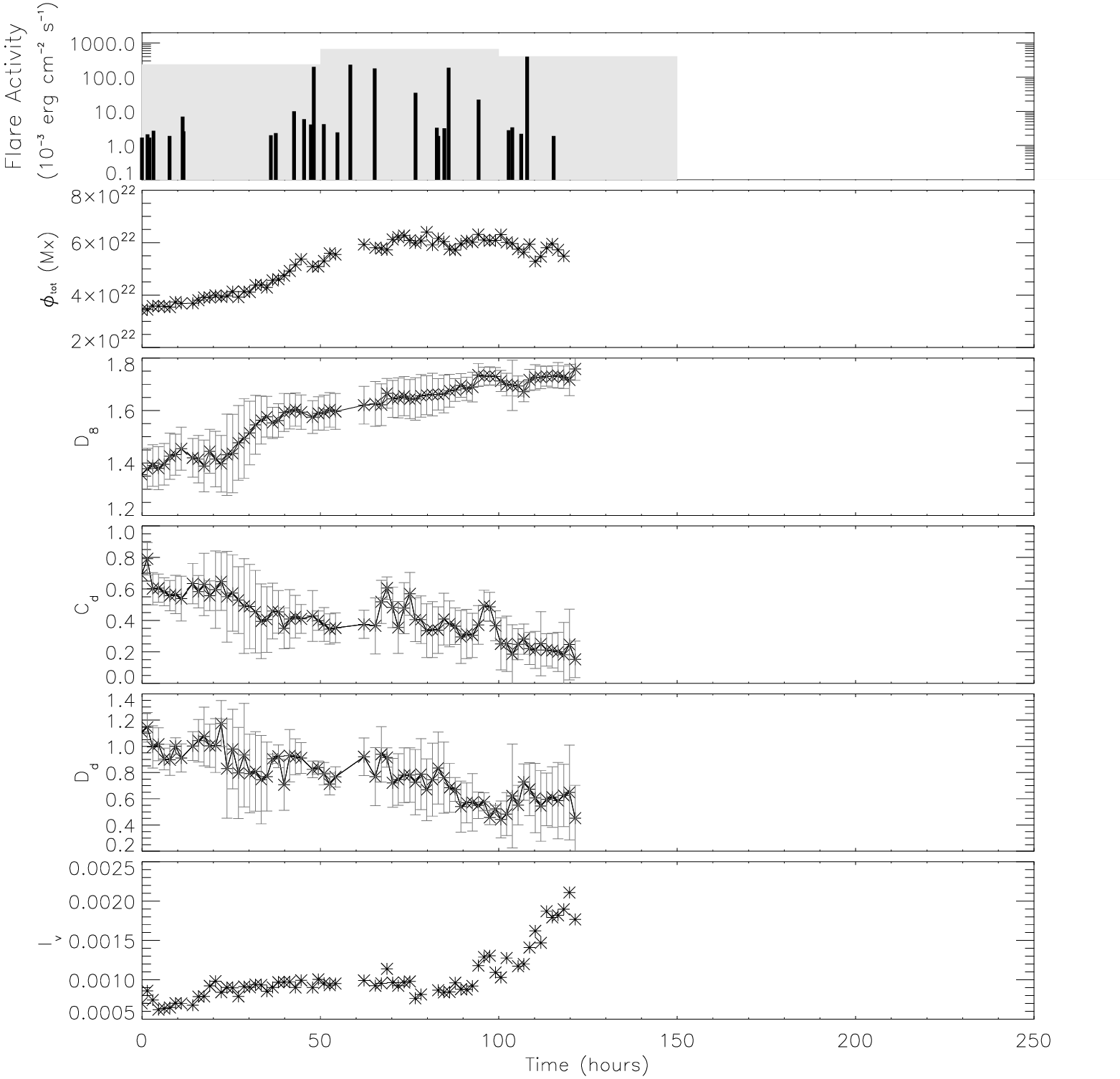}
   \caption{Flare activity, $\Phi_{tot}$, $D_8$, $C_d$, $D_d$ and $I_v$  vs. time for NOAA 9236}
   \label{Fig6}
\end{figure*}


\begin{thebibliography}{aa}%{aa}

\bibitem[Abramenko  (2005a)]{abramenko2005}
Abramenko, V.I. 2005a, \solphys{}, 228, 29 

\bibitem[Abramenko  (2005b)]{abramenko2005b}
Abramenko, V.I. 2005b, \apj{}, 629, 1141 

\bibitem[Abramenko \& Longcope (2005)]{abramenko2005c}
Abramenko, V.I. \& Longcope D.W. 2005, \apj{}, 619, 1160

\bibitem[Abramenko et al. (2008)]{abramenko2008}
Abramenko, V.I., Yurchyshyn, V., Wang, H. 2008, \apj{}, 681, 1669

\bibitem[Alissandrakis (1981)]{alissandrakis1981}
Alissandrakis, C.E. 1981, \aap{}, 100, 197 

\bibitem[Aschwanden et al. (2000)]{Asch00}
Aschwanden, M.J., Nightingale, R.W., Tarbell, T.D., Wolfson, C.J. 2000, \apj{}, 535, 1027

\bibitem[Aschwanden \& Aschwanden (2008a)]{aschwanden2008I}
Aschwanden, M.J., Aschwanden, P.D. 2008a, \apj{}, 674, 530

\bibitem[Aschwanden \& Aschwanden (2008b)]{aschwanden2008II}
Aschwanden, M.J., Aschwanden, P.D. 2008b, \apj{}, 674, 544

\bibitem[Balke et al.(1993)]{balke1993} 
Balke, A. C., Schrijver, C. J., Zwaan, C., Tarbell, T. D. 1993, \solphys{}, 143, 215

\bibitem[Barnes \& Leka (2006)]{barnes2006} 
Barnes, G., Leka, D. 2006, \apj{}, 646, 1303

\bibitem[Barnes \& Leka (2008)]{barnes2008} 
Barnes, G., Leka, D. 2008, \apj{}, 688, L107


\bibitem[B\'{e}langer et al. (2007)]{belanger2007} 
B\'{e}langer, E., Vincent, A., Charbonneau, P. 2007, \apj{}, 646, 1303

\bibitem[Bobra et al. (2008)]{bobra2008} 
Bobra, M. G., van Ballegooijen, A. A., DeLuca, E. E. 2008, \apj{}, 672, 1209

\bibitem[Bushby \& Houghton (2005)]{bushby2005} 
Bushby, P. J., Houghton, S. M. 2005, MNRAS, 362, 313

\bibitem[Cattaneo et al. (2003)]{cattaneo2003}
Cattaneo, F., Emonet, T., Weiss, N. 2003, \apj{}, 588, 1183

\bibitem[Chappell \& Scalo (2001)]{chappell2001}
Chappell, D., Scalo, J. 2001, \apj{}, 551, 712

\bibitem[Choe \& Cheng (2000)]{Choe2000}
Choe, G.S., Cheng, C.Z. 2000, \apj{}, 541, 449

\bibitem[Conlon et al. (2008)]{conlon2008}
Conlon, P.A., Gallagher, P.T., McAteer R.T.J. et al. 2008, \solphys{}, 248, 297

\bibitem[Criscuoli et al. (2007)]{criscuoli2007}
Criscuoli, S., Rast, M.P., Ermolli, I., Centrone, M. 2007, \aap{}, 461, 331 

\bibitem[Crouch et al. (2007)]{crouch2007}
Crouch, A. D., Charbonneau, P., Thibault, K. 2007 \apj{}, 662, 715

\bibitem[Freeland et al. (1998)]{freeland1998}
Freeland, S.L., Handy, B.N. 1998, \solphys{}, 182, 497 

\bibitem[Georgoulis (2005)]{georgoulis2005}
Georgoulis, M.K. 2005, \solphys{}, 228, 5

\bibitem[Georgoulis (2007)]{georgoulis2007}
Georgoulis, M.K., Rust, D.M. 2007 \apj{}, 661, 109

\bibitem[Halsey et al. (1986)]{halsey1986}
Halsey, T.C., Jensen, M.H., Kadanoff, L.P., Procaccia, I., Shraiman, B.I. 1986, \pra{}, 33, 1141 

\bibitem[Hewett et al. (2008)]{hewett2008}
Hewett, R.J., Gallagher, P.T., McAteer, R.T.J. et al. 2008, \solphys{}, 248, 311

\bibitem[Jing et al. (2009)]{jing2009}
Jing, J., Chen, P. F., Wiegelmann, T. et al. 2009, \apj{}, 696, 84

\bibitem[Kusano et al. (2002)]{kusano2002}
Kusano, K., Maeshiro, T., Yokoyama, T., Sakurai, T. 2002, \apj{}, 577, 501

\bibitem[Lawrence et al. (1995)]{lawrence1995}
Lawrence, J.K., Cadavid, A.C., Ruzmaikin, A.A. 1995, PhysRevE{}, 51, 316 

\bibitem[Lawrence et al. (1996)]{lawrence1996}
Lawrence, J.K., Cadavid, A.C., Ruzmaikin, A.A. 1996, \apj{}, 465, 425 

\bibitem[Lawrence et al. (1999)]{lawrence1999}
Lawrence, J.K., Cadavid, A.C., Ruzmaikin, A.A. 1999, \apj{}, 513, 506 

\bibitem[Leka \& Barnes (2007)]{leka2007}
Leka, K.D., Barnes, G. 2007, \apj{}, 656, 1173 
  
\bibitem[Li et al. (2004)]{li2004}
Li, Y., Luhmann, J., Fisher, G., Welsch B. 2004, J. Atm. Solar-Terrestrial Phys., 66, 1261 

\bibitem[Liu et al. (2007)]{liu2007}
Liu, Y., Norton, A.A., Scherrer, P.H. 2007, \solphys{}, 241, 185

\bibitem[McAteer et al. (2005)]{mcateer2005}
McAteer, R.T.J., Gallagher, P.T., Ireland, J. 2005, \apj{}, 631, 628

\bibitem[Longcope  (2005)]{longcope2005}
Longcope, D. 2005, LRSP, 2, 7

\bibitem[Meunier (1999)]{meunier1999} 
Meunier, N. 1999, \apj, 515, 801

\bibitem[Meunier (2004)]{meunier2004}
Meunier, N. 2004, \aap{}, 420, 333

\bibitem[Nesme-Ribes et al.(1996)]{nesme1996}
Nesme-Ribes, E., Meunier, N., \& Collin, B. 1996, \aap{}, 308, 2213

\bibitem[Parker (1988)]{Parker1988}
Parker, E.N. 1988, \apj{}, 330, 474 


\bibitem[Pietarila Graham et al. (2009)]{pietgraham2009}
Pietarila Graham, J., Danilovic, S., Sch\"{u}ssler, M. \apj{}, 693, 1972
  
\bibitem[Romano \& Zuccarello (2007)]{romano2007}
Romano, P. \& Zuccarello, F. 2007, \aap{}, 474, 633 (Paper I)

\bibitem[Schrijver et al. (1992)]{schrijver1992}
Schrijver, C. J., Zwaan, C., Balke, A. C., Tarbell, T. D., Lawrence, J. K. 1992, \aap{}, 253, 1

\bibitem[Schrijver (2007)]{schrijver2007}
Schrijver, C. J. 2007, \apj{}, 655, 117

\bibitem[Schrijver (2009)]{schrijver2009}
Schrijver, C. J. 2009, AdSpR, 43, 739

\bibitem[Song et al. (2009)]{song2009}
Song, H., Tan, C., Jing, J., Wang, H., Yurchyshyn, V., Abramenko, V. 2009, \solphys{}, 254, 101

\bibitem[Stein \& Nordlund (2006)]{stein2006}
Stein, R. F.\& Nordlund, \r{A} 2006, \aap{t}, 642, 1246 

\bibitem[Stenflo \& Holzreuter (2002)]{stenflo2002}
Stenflo, J. O. \& Holzreuter, R. 2002. In SOLMAG 2002. Proceedings of the Magnetic Coupling of the Solar Atmosphere Euroconference, ed. H. Sawaya-Lacoste, ESA SP, 505, 101

\bibitem[Tandberg-Hanssen \& Emslie  (1988)]{tandberg1988}
Tandberg-Hanssen E. \& Emslie G., The physics of solar flares, Cambridge
Astrophysics Series, 1988

\bibitem[Tao et al. (1995)]{tao1995}
Tao, L., Du, Y., Rosner, R., Cattaneo, F.1995, \apj{}, 443, 434



\bibitem[Ugarte-Urra et al. (2007)]{Ugarte2007}
Ugarte-Urra, I., Warren, H.P., Winebarger, A.R. 2007, \apj{}, 662, 1293

\bibitem[Vlahos \& Georgoulis (2004)]{vlahos2004}
Vlahos, L. \& Gergoulisa, M.K. 2004, \apj{}, 603, L61


\end{thebibliography}
\end{document}